\documentclass[twocolumn,aps,pre,amsmath,showpacs]{revtex4}

\usepackage{graphicx}

\begin{document}

\title{Connections of activated hopping processes 
with the breakdown of the Stokes-Einstein relation and 
with aspects of dynamical heterogeneities}
\author{Song-Ho Chong}
\affiliation{Institute for Molecular Science,
Okazaki 444-8585, Japan}
\date{\today}

\begin{abstract}

We develop a new extended version of the mode-coupling theory (MCT) 
for glass transition,  
which incorporates activated hopping processes
via the dynamical theory originally 
formulated to describe diffusion-jump processes in crystals.
The dynamical-theory approach adapted here to glass-forming
liquids treats hopping as arising from vibrational fluctuations in
quasi-arrested state where particles are trapped
inside their cages, and the hopping rate is formulated
in terms of the Debye-Waller factors characterizing
the structure of the quasi-arrested state. 
The resulting expression for the hopping rate
takes an activated form, and 
the barrier height for the hopping is ``self-generated'' in the sense 
that it is present only in those states where
the dynamics exhibits a well defined plateau.
It is discussed how such a hopping rate can be
incorporated into MCT so that the sharp nonergodic
transition predicted by the idealized version of the theory
is replaced by a rapid but smooth crossover.
We then show that the developed theory accounts for the breakdown 
of the Stokes-Einstein relation observed in a variety of fragile glass formers.
It is also demonstrated that characteristic features of dynamical heterogeneities
revealed by recent computer simulations are reproduced by the theory. 
More specifically, a substantial increase of the non-Gaussian parameter,
double-peak structure in the probability distribution
of particle displacements, and the presence of a growing dynamic length scale are
predicted by the extended MCT developed here, which the idealized version of the
theory failed to reproduce.
These results of the theory are demonstrated for a model of the
Lennard-Jones system, and are compared with related computer-simulation
results and experimental data. 

\end{abstract}


\maketitle

\section{Introduction}
\label{sec:introduction}

The decoupling of the self-diffusion constant from the
viscosity or the structural relaxation time
-- also referred to as the breakdown of the Stokes-Einstein (SE) relation --
that occurs for temperatures
$T \lesssim 1.2 \, T_{\rm g}$ near the
glass transition temperature $T_{\rm g}$
is among the most prominent features of
fragile glass formers~\cite{Chang97,Swallen03,Mapes06}.
The decoupling has been considered as one of the signatures
of spatially heterogeneous dynamics or 
``dynamical heterogeneities''~\cite{Ediger00,Xia01,Berthier04,Jung04}.
On the other hand, it is also well recognized that the onset
temperature $\approx 1.2 \, T_{\rm g}$ of the decoupling
is close to a crossover temperature at which 
transport properties change their characters~\cite{Roessler90}
and below which the dynamics is thought to be dominated by 
activated hopping processes over barriers~\cite{Goldstein69}.
Then, a natural question arises as to possible connections 
among the decoupling, dynamical heterogeneities, and the 
hopping processes. 
In this paper, such connections are explored by extending
the idealized mode-coupling theory (MCT) for 
glass transition~\cite{Goetze91b}. 

The idealized MCT has been known as the most successful microscopic theory
for glass transition.
Indeed, extensive tests of the theoretical predictions carried out so far
against experimental data and computer-simulation results
suggest that the theory deals properly with some essential features of glass-forming
liquids~\cite{Goetze92,Goetze99}.
On the other hand, a well-recognized limitation of the
idealized MCT is the predicted divergence of the
$\alpha$-relaxation time at a critical temperature $T_{\rm c}$
-- also referred to as the nonergodic transition --
which is not observed in experiments and computer simulations. 
An extended version of MCT developed in Ref.~\cite{Goetze87} aims at
incorporating activated hopping processes which smear out the sharp
nonergodic transition and restore ergodicity for $T \le T_{\rm c}$,
but its applicability has been restricted to schematic models.
This is because of the presence of the subtraction term in the expression
for the hopping kernel, which violates the positiveness
-- a fundamental property -- of any correlation spectrum. 

There have been relatively few other attempts to go beyond
the idealized MCT~\cite{Kawasaki94,Schweizer03,Bhattacharyya05},
and incorporating the hopping processes into the theory for glass transition
has been a major unsolved problem.
We present here a new formulation
which is motivated by ideas from the dynamical theory
originally developed to describe 
diffusion-jump processes in crystals~\cite{Flynn-all}.
The dynamical-theory approach adapted in this work to glass-forming
liquids treats hopping as arising from vibrational fluctuations in
quasi-arrested state where particles are trapped
inside their cages, and the hopping rate is formulated
in terms of the Debye-Waller factors characterizing
the structure of the quasi-arrested state. 
The resulting expression for the hopping rate takes an activated form, and 
the barrier height for the hopping is ``self-generated'' in the sense 
that it is present only in those states where
the dynamics exhibits a well defined plateau.
It will be discussed how such a hopping rate can be
incorporated to develop a new extended version of MCT. 

We will then investigate whether the developed 
theory accounts for the mentioned decoupling for $T \lesssim T_{\rm c}$.
Such an investigation makes sense 
since $T_{\rm c}$ is also found to be close to 
$1.2 \, T_{\rm g}$~\cite{Roessler90},
i.e., the onset temperature of the decoupling~\cite{Chang97,Zollmer03}.
It will also be examined whether our theory reproduces characteristic
features of dynamical heterogeneities revealed by recent computer simulations.
More specifically, we shall study whether the theory predicts
a substantial increase of the non-Gaussian parameter~\cite{Kob97},
double-peak structure in the probability distribution of particle 
displacements~\cite{Flenner05,Flenner05b},
and the presence of a growing dynamic length scale~\cite{Berthier04}, 
which the idealized MCT failed to reproduce.

The paper is organized as follows.
In Sec.~\ref{sec:theory}, we formulate our new extended MCT. 
Numerical results of the theory are presented in Sec.~\ref{sec:results}
for a model of the Lennard-Jones system, and
connections of the hopping processes with the breakdown of the
SE relation and with aspects of dynamical heterogeneities are discussed.
The paper is summarized in Sec.~\ref{sec:conclusion}.
Appendix~\ref{appendix:GS-theory} outlines a novel derivation
of the hopping kernel formulated in Ref.~\cite{Goetze87}, and
Appendix~\ref{appendix:MSD-NGP} is devoted to 
the derivation of the extended-MCT equations for the
mean-squared displacement and the non-Gaussian parameter.

\section{Theory}
\label{sec:theory}

We start from surveying basic features of 
the idealized MCT~\cite{Goetze91b} (see also Appendix~\ref{appendix:GS-theory}).
A system of $N$ atoms of mass $M$ distributed with
density $\rho$ shall be considered. 
Structural changes as a function of time $t$ are characterized
by coherent density correlators
$\phi_{q}(t) = \langle 
  \rho_{\vec q}^{*} \, e^{i {\cal L} t} \rho_{\vec q}
\rangle / NS_{q}$.
Here $\rho_{\vec q} = 
\sum_{i} \exp( i {\vec q} \cdot {\vec r}_{i})$
with ${\vec r}_{i}$ referring to $i$th particle's position
denotes density fluctuations for wave vector ${\vec q}$;
${\cal L}$ the Liouville operator; 
$\langle \cdot \rangle$ the canonical averaging
for temperature $T$;
$S_{q} = \langle \rho_{\vec q}^{*} \, \rho_{\vec q} \rangle / N$
the static structure factor; and
$q = | \, {\vec q} \, |$. 
Within the Zwanzig-Mori formalism~\cite{Hansen86}
one obtains the following exact equation of motion:
\begin{subequations}
\label{eq:GLE-phi}
\begin{equation}
\partial_{t}^{2} \phi_{q}(t) +
\Omega_{q}^{2} \phi_{q}(t) +
\Omega_{q}^{2} \int_{0}^{t} dt^{\prime} \,
m_{q}(t-t^{\prime}) \, \partial_{t^{\prime}} \phi_{q}(t^{\prime}) = 0.
\label{eq:GLE-phi-a}
\end{equation}
Here $\Omega_{q}^{2} = q^{2} k_{\rm B}T/MS_{q}$ 
with Boltzmann's constant $k_{\rm B}$, and
the memory kernel $m_{q}(t)$ describes correlations of fluctuating forces.
Introducing the Laplace transform with the convention
$f(z) = i \int_{0}^{\infty} dt \, e^{izt} f(t)$ (${\rm Im} \, z > 0$),
Eq.~(\ref{eq:GLE-phi-a}) is equivalent to the representation
\begin{equation}
\phi_{q}(z) = 
- 1 \, / \, 
  \{ z - \Omega_{q}^{2} \, / \, 
    [z + \Omega_{q}^{2} \, m_{q}(z)] \}.
\label{eq:GLE-phi-b}
\end{equation}
\end{subequations}
Under the mode-coupling approximation,
the fluctuating forces are approximated by their projections
onto the subspace spanned by pair-density modes
$\rho_{\vec k} \rho_{\vec p}$.
The factorization approximation for dynamics
of the pair-density modes yields the following idealized-MCT
expression for the memory kernel to be denoted as
$m_{q}^{\rm id}(t)$:
\begin{subequations}
\label{eq:ideal-MCT-m}
\begin{equation}
m_{q}^{\rm id}(t) =
\int d{\vec k} \,
V({\vec q}; {\vec k}, {\vec p} \, ) \,
\phi_{k}(t) \, \phi_{p}(t).
\label{eq:ideal-MCT-m-a}
\end{equation}
Here ${\vec p} = {\vec q} - {\vec k}$, and the vertex function is given by
\begin{equation}
V({\vec q}; {\vec k}, {\vec p} \, ) =
\frac{\rho}{2(2 \pi)^{3} q^{4}} S_{q} S_{k} S_{p}
[ ({\vec q} \cdot {\vec k}) c_{k} + ({\vec q} \cdot {\vec p}) c_{p} ]^{2},
\end{equation}
\end{subequations}
in terms of $S_{q}$ and the direct correlation function
$c_{q} = (1 - 1/S_{q}) / \rho$.
The idealized-MCT equations (\ref{eq:GLE-phi})
and (\ref{eq:ideal-MCT-m}) 
exhibit a bifurcation for 
$\phi_{q}(t \to \infty) = f_{q}$ --
also referred to as the nonergodic transition --
at a critical temperature $T_{\rm c}$~\cite{Goetze91b}. 
For $T > T_{\rm c}$, 
the correlator relaxes towards 
$f_{q} = 0$ as expected for ergodic liquid states. 
On the other hand,
density fluctuations for $T \le T_{\rm c}$
arrest in a disordered solid,
quantified by a Debye-Waller factor $f_{q} > 0$.

It is the factorization approximation [{\em cf.} Eq.~(\ref{eq:dynamic-factorization})]
that leads to the nonergodic transition at $T_{\rm c}$. 
Therefore, one has to consider corrections,
$m_{q}(z) = m_{q}^{\rm id}(z) + \Delta m_{q}(z)$, 
to go beyond the idealized MCT, 
which shall be quantified via the hopping kernel
defined by 
\begin{subequations}
\label{eq:extended-memory}
\begin{equation}
\delta_{q}(z) = - 1/m_{q}(z) + 1/m_{q}^{\rm id}(z).
\label{eq:extended-memory-a}
\end{equation}
The correction term reads
$\Delta m_{q}(z) = m_{q}^{\rm id}(z) \delta_{q}(z) m_{q}(z)$,
and the memory kernel $m_{q}(z)$ can be expressed as
\begin{equation}
m_{q}(z) = m_{q}^{\rm id}(z) \, / \,
[ \, 1 - \delta_{q}(z) \, m_{q}^{\rm id}(z) \, ].
\label{eq:extended-memory-b}
\end{equation}
\end{subequations}
As demonstrated in Appendix~\ref{appendix:GS-theory},
one can derive based on Eq.~(\ref{eq:extended-memory-a})
an expression for $\delta_{q}(z)$ which
is essentially the same as that of 
the extended MCT of G\"otze and Sj\"ogren~\cite{Goetze87,Goetze95}
by applying the Zwanzig-Mori formalism to $m_{q}(t)$ and then
introducing the corresponding mode-coupling approximation
for the ``memory kernel'' to $m_{q}(t)$.
Such a hopping kernel, however, retains the same problem
mentioned in Sec.~\ref{sec:introduction}. 
Instead, our attempt for the extension of the idealized MCT
is motivated by the following observation:
substituting Eq.~(\ref{eq:extended-memory-b}) into
Eq.~(\ref{eq:GLE-phi-b}) yields
for small $z$~\cite{Goetze92} 
\begin{equation}
\phi_{q}(z) = 
- 1 \, / \, 
  \{ z + \delta_{q}(z) - \Omega_{q}^{2} \, / \, 
    [z + \Omega_{q}^{2} \, m_{q}^{\rm id}(z)] \}.
\label{eq:alternative-GLE}
\end{equation}
Dropping $\delta_{q}(z)$, this equation reduces to the
one of the idealized MCT: 
approaching $T_{\rm c}$ from above, 
$m_{q}^{\rm id}(z)$ for small $z$ becomes larger, 
and so does $\phi_{q}(z)$, 
leading to the nonergodic transition at $T = T_{\rm c}$. 
In the presence of $\delta_{q}(z)$, on the other hand,
the transition is cutoff since 
the third term in the denominator of Eq.~(\ref{eq:alternative-GLE})
becomes unimportant when
$m_{q}^{\rm id}(z)$ becomes large.
The long-time dynamics of $\phi_{q}(t)$ in this case is 
thus determined by $\delta_{q}(z)$ for small $z$. 
This observation raises a possibility of constructing a
new approximate theory for $\delta_{q}(z)$ described below. 

We first derive a rate formula for a hopping process in which
an atom at ${\vec r}_{i}$ jumps to a
nearby site ${\vec r}_{i}^{\, \prime}$ separated by an
interparticle distance.
The presence of such a process at low temperatures 
has been revealed by computer 
simulations~\cite{Roux89,Wahnstroem91,Schroder00}. 
This will be done via the dynamical theory
originally developed
to describe diffusion-jump processes in 
crystals~\cite{Flynn-all}.
The approach adapted here to glass-forming liquids
treats hopping as arising from vibrational fluctuations (phonons) 
in quasi-arrested state where particles are trapped
inside their cages.
Let us suppose that the quasi-arrested state
characterized by the Debye-Waller factors $f_{q}$  
can be described as a frozen, irregular lattice~\cite{Barrat89}.
Each particle then has a well defined equilibrium
position ${\vec R}_{i}$ within the lifetime of the quasi-arrested state, and we
introduce the displacement
from the equilibrium position via
${\vec r}_{i} = {\vec R}_{i} + {\vec u}({\vec R}_{i})$.
The essential feature of the hopping process 
is that a jumping atom 
passes over a barrier formed by neighbors 
which block a direct passage to the new site. 
The criterion that determines whether or not a given
fluctuation is sufficient to cause
a jump is therefore concerned with the relative
displacements of the atom and the saddlepoint.
We thus employ as a ``reaction coordinate''~\cite{Flynn-all},
$x(t) = [
{\vec u}({\vec R}_{i}+{\vec s},t) - {\vec u}({\vec R}_{i},t) ]
\cdot \hat{\vec s}$,
and assume that a hopping occurs when 
$x(t)$ exceeds a critical value $x^{*}$, which measures the 
size of fluctuation needed to cause a jump.
Here ${\vec s}$ denotes the saddlepoint
position with respect to ${\vec R}_{i}$, and
the scalar product selects only those fluctuations 
directed towards $\hat{\vec s} = {\vec s} / s$.
Each phonon displaces a hopping atom towards the saddlepoint.
The phonon phases are random, but
the displacements may
occasionally coincide in such a way that a hopping process
occurs.
The hopping rate $w_{\rm hop}$ can then be calculated from 
such a probability per unit time,
and one obtains along the line described in Ref.~\cite{Flynn-all}
with the isotropic Debye approximation, 
$w_{\rm hop} = (1/2\pi) (3/5)^{1/2} \omega_{\rm D}
\exp[- 3mv^{2} \Delta^{2} / 2 k_{\rm B}T ]$,
in terms of the sound velocity $v$. 
Here $\omega_{\rm D} = k_{\rm D} v$ with the Debye wave number
$k_{\rm D} = (6 \pi^{2} \rho)^{1/3}$,
and $\Delta \equiv x^{*} / s$. 
Notice that the sound velocity here refers to
the one in the quasi-arrested state, 
which is renormalized by the Debye-Waller 
factors~\cite{Goetze00,Chong06}.
To emphasize this, the sound velocity shall be expressed as
$v = \sqrt{M_{\rm L}/(\rho m)}$
in terms of the (longitudinal) elastic modulus
\begin{subequations}
\begin{equation}
M_{\rm L} = M_{\rm L}^{0} + \delta M_{\rm L},
\end{equation}
consisting of the equilibrium value 
$M_{\rm L}^{0} = \rho (k_{\rm B}T) / S_{0}$,
where $S_{0} \equiv S_{q \to 0}$, and
an additional contribution for the quasi-arrested state, 
for which MCT yields~\cite{Goetze03}
\begin{equation}
\delta M_{\rm L} =  \rho (k_{\rm B}T)
\int dk \, V_{\rm L}(k) f_{k}^{2},
\end{equation}
with 
\begin{equation}
V_{\rm L}(k) = 
\frac{\rho k^{2} S_{k}^{2}}{4 \pi^{2}}
\left\{
  c_{k}^{2} + \frac{2}{3} [ k c_{k}^{\prime} ] c_{k} + 
  \frac{1}{5} [ k c_{k}^{\prime}]^{2}
\right\}.
\end{equation}
\end{subequations}
The hopping rate is then given by 
\begin{equation}
w_{\rm hop} = \frac{1}{2 \pi} 
\left( \frac{3}{5} \right)^{\frac{1}{2}} 
\omega_{\rm D} 
\exp 
\left[
  - \frac{3 M_{\rm L}}{2 \rho k_{\rm B}T} \, \Delta^{2}
\right].
\label{eq:w-Kac-Debye-2}
\end{equation}
The pre-exponential factor 
represents a mean attack frequency while the 
exponential term of the activated form 
gives the probability that the system is
found at the critical displacement $x^{*}$.
In addition, the barrier height for the hopping is
``self-generated'' in the sense that it is determined by 
the plateau heights $f_{q}$ of the coherent density correlators,
and is present only in those states where
the dynamics exhibits a well defined plateau.

We next relate the hopping rate $w_{\rm hop}$
to the hopping kernel $\delta_{q}(z)$. 
Our discussion becomes simpler if 
the tagged-particle density correlator
$\phi_{q}^{s}(t)$ 
-- the self part of $\phi_{q}(t)$ -- 
is considered, 
so this case shall be considered first.
Hereafter, quantities referring to the tagged particle shall
be marked with the superscript or subscript ``$s$''.
The Zwanzig-Mori equation for $\phi_{q}^{s}(t)$
has the same form as Eq.~(\ref{eq:GLE-phi-a})
with $\phi_{q}$, $m_{q}$, and $\Omega_{q}^{2}$ replaced by
$\phi_{q}^{s}$, $m_{q}^{s}$, and
$(\Omega_{q}^{s})^{2} = q^{2} k_{\rm B}T / M$, respectively;
the idealized-MCT kernel corresponding to Eq.~(\ref{eq:ideal-MCT-m-a})
is given by
$m_{q}^{s \, {\rm id}}(t) = 
\int d{\vec k} \, V^{s}({\vec q}; {\vec k}, {\vec p}) \phi_{k}(t) \phi_{p}^{s}(t)$
with 
$V^{s} =  \rho S_{k} [ ({\vec q} \cdot {\vec k}) c_{k} ] ^{2} / [ (2 \pi)^{3} q^{4} ]$~\cite{Fuchs98};
and Eq.~(\ref{eq:extended-memory-b}) holds with
$m_{q}$, $m_{q}^{\rm id}$, and $\delta_{q}$ replaced by
$m_{q}^{s}$, $m_{q}^{s \, {\rm is}}$, and $\delta_{q}^{s}$, respectively. 
The arrested part $f_{q}^{s}$ of the correlator
$\phi_{q}^{s}(t)$ is referred to as the Lamb-M\"ossbauer factor.

In the absence of the hopping kernel, the idealized
kernel $m_{q}^{s \, {\rm id}}(t)$ for $T \le T_{\rm c}$ 
arrests at a plateau for long times
whose height is given by 
$C_{q}^{s} = f_{q}^{s} / (1 - f_{q}^{s})$~\cite{Goetze91b},
i.e.,
there holds 
$m_{q}^{s \, {\rm id}}(z) = - C_{q}^{s} / z$ for small $z$.
Substituting this into Eq.~(\ref{eq:alternative-GLE}) for $\phi_{q}^{\rm s}(z)$ 
yields for small $z$
\begin{equation}
\phi_{q}^{s}(z) = - f_{q}^{s} / [ z + f_{q}^{s} \delta_{q}^{s}(z)],
\label{eq:phis-small-z-1}
\end{equation}
which determines the $\alpha$ relaxation
of $\phi_{q}^{s}(t)$ -- the decay from the plateau $f_{q}^{s}$ to zero --
in the presence of $\delta_{q}^{s}(z)$. 

On the other hand, when the $\alpha$ relaxation is 
dominated by hopping processes characterized by a rate 
$w_{\rm hop}({\vec r} \to {\vec r}^{\, \prime})$,
the van Hove self correlation function $G_{s}({\vec r},t)$,
related to $\phi_{q}^{s}(t)$ via the inverse Fourier transform
\begin{equation}
G_{s}({\vec r}, t) =
\frac{1}{(2 \pi)^{3}}
\int d {\vec q} \, e^{- i {\vec q} \cdot {\vec r}} \,
\phi_{q}^{s}(t),
\end{equation}
and proportional to the probability 
of finding the tagged particle at ${\vec r}$ and $t$~\cite{Hansen86}, 
obeys a simple rate equation
\begin{eqnarray}
\partial_{t} G_{s}({\vec r},t) &=&
\sum_{\vec l}
[ \, w_{\rm hop}({\vec r} + {\vec l} \to {\vec r})
G_{s}({\vec r} + {\vec l},t) 
\nonumber \\
& & \qquad \quad
- \,
w_{\rm hop}({\vec r} \to {\vec r}+{\vec l})
G_{s}({\vec r},t) \, ].
\end{eqnarray}
We assume that only hoppings with $| \, {\vec l} \, | \approx a$ 
are relevant, in which
$a \equiv \int_{0}^{r_{\rm min}} dr \, r \left[ N(r)/N_{\rm c} \right]$
denotes the weighted average 
of interparticle distances.
Here $r_{\rm min}$ denotes the first minimum of the
radial distribution function $g(r)$ defining the first shell;
$N(r)dr$ with $N(r)  = 4 \pi r^{2} \rho g(r)$ 
gives the mean number of particles at distance
between $r$ and $r + dr$;
and $N_{\rm c} = \int_{0}^{r_{\rm min}} dr N(r)$ is 
the coordination number of the first shell. 
The quantity $a$ serves as an analogue of the lattice spacing in crystals. 
Since there is no site dependence in the
hopping rate we formulated [{\em cf.} Eq.~(\ref{eq:w-Kac-Debye-2})],
there holds
\begin{equation}
\partial_{t} G_{s}({\vec r},t) =
w_{\rm hop} 
\sum_{|{\vec l}| \approx a}
[ \, G_{s}({\vec r} + {\vec l},t) - G_{s}({\vec r},t) \, ].
\end{equation} 
Fourier transforming this yields
\begin{equation}
\partial_{t} \phi_{q}^{s}(t) =
- w_{\rm hop} 
\sum_{| {\vec l} | \approx a}
[ \, 1 - e^{- i {\vec q} \cdot {\vec l}} \, ] \,
\phi_{q}^{s}(t).
\end{equation}
Assuming that ${\vec l}$ are oriented at random,
the summation $\sum_{| \, {\vec l} \, | \approx a}$
is given by the orientational average multiplied by the
number of sites satisfying $| \, {\vec l} \, | \approx a$,
which is approximated by the coordination number $N_{\rm c}$
of the first shell. 
This leads to
\begin{equation}
\partial_{t} \phi_{q}^{s}(t) = - 
w_{\rm hop} N_{\rm c} [1 - \sin(qa)/(qa)] \phi_{q}^{s}(t).
\end{equation}
Noticing that the $\alpha$ relaxation of $\phi_{q}^{s}(t)$ starts
from the plateau $f_{q}^{s}$, the Laplace transform of this equation reads
\begin{equation}
\phi_{q}^{s}(z) = - 
f_{q}^{s} \, / \, \{ \, z + i w_{\rm hop} N_{\rm c} [1 - \sin(qa)/(qa)]  \, \}.
\label{eq:phis-small-z-2}
\end{equation}
By comparing Eqs.~(\ref{eq:phis-small-z-1}) and (\ref{eq:phis-small-z-2}),
one arrives at the following expression for the hopping kernel:
\begin{equation}
\delta_{q}^{s}(z) = 
i \, w_{\rm hop} N_{\rm c} [1 - \sin(qa)/(qa)] / f_{q}^{s}.
\label{eq:delta-self}
\end{equation}

The collective hopping kernel $\delta_{q}(z)$
consists of the self and distinct parts, 
$\delta_{q}(z) = \delta_{q}^{s}(z)/S_{q} +
\delta_{q}^{\rm dist}(z)$.
In the present work, we shall adopt a simple
model for $\delta_{q}(z)$ 
in which the distinct part describing 
possible correlated jumps is neglected: 
\begin{equation}
\delta_{q}(z) = \delta_{q}^{s}(z) / S_{q}.
\label{eq:delta-collective}
\end{equation}
This model for $\delta_{q}(z)$ looks oversimplified, but
nontrivial theoretical predictions follow from such a simple
model as will be demonstrated in Sec.~\ref{sec:results}.

Equations~(\ref{eq:GLE-phi-a}) and
(\ref{eq:extended-memory-b}) with
Eqs.~(\ref{eq:ideal-MCT-m-a}), 
(\ref{eq:w-Kac-Debye-2}),
(\ref{eq:delta-self}), and (\ref{eq:delta-collective})
constitute our new extended-MCT equations 
for the coherent density correlator $\phi_{q}(t)$;
corresponding equations hold for the tagged-particle
density correlator $\phi_{q}^{s}(t)$ with
the aforementioned replacement of 
$\phi_{q}$, $m_{q}$, $\Omega_{q}^{2}$, $m_{q}^{\rm id}$, and $\delta_{q}$
by
$\phi_{q}^{s}$, $m_{q}^{s}$, $(\Omega_{q}^{s})^{2}$, 
$m_{q}^{s \, {\rm id}}$, and $\delta_{q}^{s}$,
respectively. 
The extended-MCT equations for the mean-squared displacement 
and the non-Gaussian parameter, which are required for our discussion
in Sec.~\ref{sec:results}, 
are derived in Appendix~\ref{appendix:MSD-NGP}. 
All these equations can be solved provided
$S_{q}$ and $\Delta^{2}$
are known as input.
(We notice that $f_{q}$ and $f_{q}^{s}$
can be obtained based on the knowledge of
$S_{q}$~\cite{Goetze91b}.)

\section{Results and discussion}
\label{sec:results}

In the following, numerical results of the 
extended theory will be presented for the
Lennard-Jones (LJ) system in which
particles interact via the potential
$V(r) = 4 \epsilon_{\rm LJ} 
\{ (\sigma_{\rm LJ}/r)^{12} - (\sigma_{\rm LJ}/r)^{6} \}$.
$S_{q}$ shall be evaluated within the
Percus-Yevick approximation~\cite{Hansen86}.
This model has been studied in Ref.~\cite{Chong06} based on the
idealized MCT. 
From here on, all quantities are expressed in reduced units
with the unit of length $\sigma_{\rm LJ}$, the unit of energy 
$\epsilon_{\rm LJ}$ (setting $k_{\rm B} = 1$),
and the unit of time $(m\sigma_{\rm LJ}^{2}/\epsilon_{\rm LJ})^{1/2}$.
The dynamics as a function of $T$ shall be considered for 
a fixed density $\rho = 1.093$, for which
the critical temperature of the idealized MCT is found to be
$T_{\rm c} \approx 1.637$~\cite{Chong06}.
For $\Delta^{2}$ entering into the extended-MCT equations, 
we set $\Delta^{2} = 0.10$ estimated in Ref.~\cite{Flynn-all}
from migration properties of crystals.
We notice that this value of $\Delta^{2}$ 
is consistent with the Lindemann length~\cite{comment-Lindemann-length}.
We also confirmed that the results to be presented below do not
rely on the specific value of $\Delta^{2} = 0.10$:
nearly the same results were obtained with other values of $\Delta^{2}$,
as far as those values 
consistent with the Lindemann length are chosen. 

\subsection{Coherent density correlators}

\begin{figure}[tb]
\centerline{\includegraphics[width=0.8\linewidth]{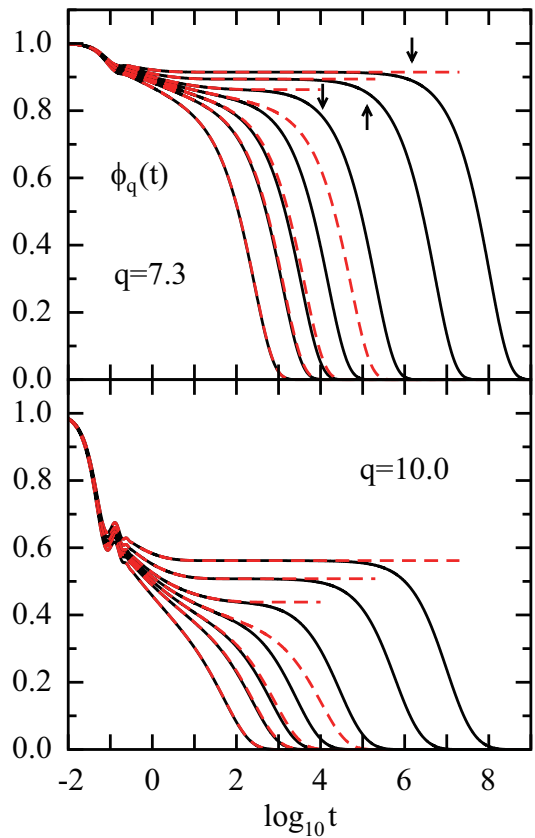}}
\caption{
(Color online)
Coherent density correlators $\phi_{q}(t)$ as a function of
$\log_{10}t$
for reduced temperatures
$\epsilon \equiv (T_{\rm c} - T)/T_{\rm c} = -0.10$,
$-0.05$, $-0.03$, $-0.01$, $+0.01$, 
$+0.05$, and $+0.10$
(from left to right).
The wave numbers are $q = 7.3$ (upper panel) and
$q = 10.0$ (lower panel), which 
correspond to the first-peak and first-minimum 
positions of $S_{q}$, respectively. 
The solid and dashed curves denote the results from 
the extended and idealized MCT,
respectively.
The arrows in the upper panel refer to the peak positions of the
non-Gaussian parameter $\alpha_{2}(t)$ 
for $\epsilon = +0.01$, $+0.05$, and $+0.10$ 
({\em cf.} Sec.~\ref{subsec:DH}).}
\label{fig:phi}
\end{figure}

Figure~\ref{fig:phi} shows the coherent density correlators
$\phi_{q}(t)$ for representative reduced temperatures 
$\epsilon \equiv (T_{\rm c}-T)/T_{\rm c}$
whose values are specified in the caption. 
The wave numbers shown are $q = 7.3$ (upper panel) and
$q = 10.0$ (lower panel), which 
correspond to the first-peak and first-minimum 
positions of $S_{q}$, respectively. 
The dashed curves refer to the idealized-MCT results which exhibit the
ergodic ($\epsilon < 0$) to nonergodic ($\epsilon \ge 0$)
transition at $T = T_{\rm c}$ ($\epsilon = 0$)~\cite{Goetze91b,Franosch97}.
The solid curves denote the results from the extended MCT.

It is seen from Fig.~\ref{fig:phi} 
that the solid curves for $\epsilon = -0.10$ and $-0.05$ 
are hardly affected by the hopping processes, but 
a slight deviation from the dashed curve is discernible in the
$\alpha$-relaxation regime of the solid curve
for $\epsilon = -0.03$.
The solid curve for $\epsilon = -0.01$ exhibits the same decay 
up to $\log_{10} t \approx 2$ 
as the corresponding dashed curve, 
but the relaxation thereafter is considerably accelerated. 
The effects from the hopping processes are
drastic for $\epsilon > 0$ where 
the idealized MCT predicts the arrested dynamics
at the plateau $f_{q}$, 
whereas the corresponding solid curves from 
the extended theory relax to zero for long times.

It would be interesting to analyze these extended-MCT results 
based on various scaling laws developed in Ref.~\cite{Goetze87}.
Such an analysis, however, shall be deferred to subsequent publications,
and in the following, 
we will focus on the connections of the hopping processes with the
breakdown of the SE relation and with aspects of dynamical heterogeneities. 

\subsection{Breakdown of the Stokes-Einstein relation}
\label{subsec:breakdown}

Here 
we investigate the breakdown of the SE relation based on the
extended MCT, and compare our theoretical prediction with 
related computer-simulation results and experimental data.
A convenient experimental measure of the breakdown is a product
$D \eta/T$ formed with the diffusion coefficient $D$ and the viscosity $\eta$,
which grows as the SE relation fails~\cite{Chang97}.
In the present study, the $\alpha$-relaxation time $\tau_{q^{*}}$ 
of the coherent density correlator 
at the peak position $q^{*} = 7.3$ of $S_{q}$, 
defined via the convention $\phi_{q^{*}}(\tau_{q^{*}}) = 0.1$, 
shall be used as a substitute for $\eta/T$.
This is justified since the $T$ dependence of the $\alpha$-relaxation time
at the structure factor peak 
is known to track that of $\eta/T$~\cite{Mezei87,Yamamoto98}.
The diffusion coefficient is determined from the long-time asymptote
$D = \lim_{t \to \infty} \delta r^{2}(t) / 6t$
of the mean-squared displacement
$\delta r^{2}(t) \equiv
\langle [ {\vec r}_{s}(t) - {\vec r}_{s}(0) ]^{2} \rangle$,
whose extended-MCT equations are derived in 
Appendix~\ref{appendix:MSD}.

\begin{figure}[tb]
\includegraphics[width=1.0\linewidth]{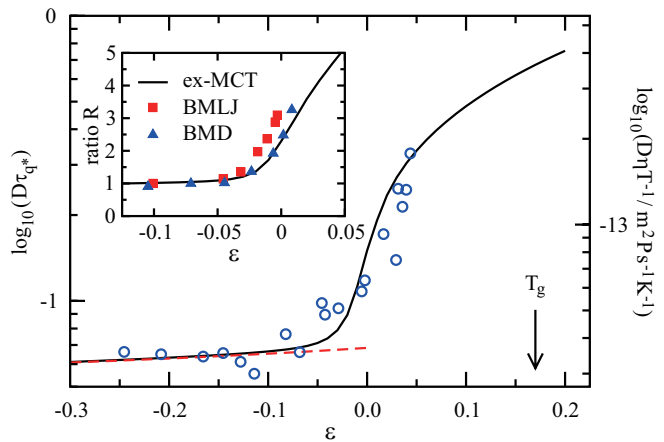}
\caption{
(Color online)
Logarithmic representation of the product $D \tau_{q^{*}}$ 
as a function of the reduced temperature
$\epsilon = (T_{\rm c} - T)/T_{\rm c}$.
The solid and dashed curves, both referring to the left scale,
denote the results from the extended and idealized MCT,
respectively.
Circles (right scale) denote 
the $D \eta / T$ data of salol taken from Ref.~\cite{Chang97} and
plotted versus $\epsilon$ with $T_{\rm c} = 262.7$ K~\cite{Li92b}.
The arrow marks $T_{\rm g} = 218$ K of salol.
Both the left and right ordinates range over 1.3 decades.
The inset exhibits the ratio $R$ of the product $D \tau_{q^{*}}$ to the one at a
reference temperature where the SE relation holds.
The prediction from the extended MCT (solid curve) is compared with 
simulation results for a binary mixture of Lennard-Jones 
particles~\cite{Kob94} (filled squares) and 
a binary mixture of dumbbell molecules of elongation 
$\zeta = 0.8$~\cite{Chong05} (filled triangles).
The procedure for the comparison is detailed in 
Ref.~\cite{comment-comparison-with-simulation}.}
\label{fig:Dtau}
\end{figure}

Figure~\ref{fig:Dtau} shows the theoretical prediction for 
the product $D \tau_{q^{*}}$ as
a function of the reduced temperature 
$\epsilon = (T_{\rm c} - T)/T_{\rm c}$.
The solid and dashed curves, both referring to the left scale,
denote the results from the extended and idealized MCT, 
respectively.
The idealized-MCT result for $D \tau_{q^{*}}$
varies little for $\epsilon < 0$
(the increase is only about 10\% for the $\epsilon$ range
shown in Fig.~\ref{fig:Dtau}), and does not account
for the breakdown of the SE relation.
This reflects the universal $\alpha$-scale coupling predicted by the
idealized MCT~\cite{Goetze91b}, according to which
both the $\alpha$-relaxation time $\tau_{q^{*}}$ and 
the inverse of the diffusivity $1/D$
exhibit a universal power-law behavior $| \, \epsilon \, |^{-\gamma}$ 
for small $\epsilon$, and hence,  
the product $D \tau_{q^{*}}$ approaches a constant for
$\epsilon \to 0-$ ($T \to T_{\rm c}+$). 
On the other hand,
the extended MCT predicts the increase of the product
$D \tau_{q^{*}}$ for $\epsilon \gtrsim -0.05$.
Figure~\ref{fig:Dtau} thus shows one of the main
results of this paper that the
hopping processes are responsible for the breakdown of
the SE relation.

Also presented in Fig.~\ref{fig:Dtau} is the comparison of the
theoretical result with the $D \eta / T$ data of salol
taken from Ref.~\cite{Chang97} (circles, right scale).
The experimental data are also plotted versus $\epsilon$
with $T_{\rm c} = 262.7$~K
of salol determined in Ref.~\cite{Li92b}.
A direct comparison between the theoretical result and the
experimental data can be made by plotting them on logarithmic
scales of the same range as done in Fig.~\ref{fig:Dtau}. 
It is seen that the extended-MCT result is consistent with
the experimental data concerning the degree of the
breakdown of the SE relation.

The inset of Fig.~\ref{fig:Dtau} compares the extended-MCT
result (solid curve) with simulation results
for a binary mixture of Lennard-Jones particles~\cite{Kob94}
(filled squares) and a binary mixture of
dumbbell molecules of elongation $\zeta = 0.8$~\cite{Chong05}
(filled triangles).
Here the comparison is done in terms of the ratio $R$ 
of the product $D \tau_{q^{*}}$ to the one at a reference temperature
where the SE relation holds.
(See Ref.~\cite{comment-comparison-with-simulation}
concerning the details of the comparison.)
By definition, the ratio $R$ is unity if the SE relation holds, 
whereas it exceeds unity as the SE relation fails.
Again, the theoretical prediction is consistent with the simulation results.

On the other hand, 
recent measurements~\cite{Swallen03,Mapes06} indicate that
the self-diffusion coefficient is about 100 times faster
near $T_{\rm g}$ than that predicted by 
the SE relation.
This is about a factor of 10 larger compared to our 
theoretical prediction near $T_{\rm g}$ ({\em cf.} Fig.~\ref{fig:Dtau}), 
assuming that $T_{\rm g}$ of the present system is located 
at $\epsilon \approx 0.17$ estimated from the
reduced temperature at $T_{\rm g}$ of salol.
This implies that our model for the hopping kernel
might be too primitive to be applicable near $T_{\rm g}$.
In the following, we shall therefore focus mainly on the regime 
$\epsilon \lesssim +0.05$ where our theoretical
prediction is consistent with the simulation results and experimental data.

\subsection{Aspects of dynamical heterogeneities}
\label{subsec:DH}

\subsubsection{Non-Gaussian parameter}

We next explore a connection of the hopping processes with
aspects of dynamical heterogeneities. 
We start from the discussion on the non-Gaussian parameter
$\alpha_{2}(t)$ which characterizes deviations from the
Gaussian behavior of the van Hove self correlation function~\cite{Rahman64}:
\begin{equation}
\alpha_{2}(t) \equiv \frac{3}{5}
\left[ \delta r^{4}(t) / \delta r^{2}(t)^{2} \right] - 1.
\label{eq:NGP-def}
\end{equation}
Here 
$\delta r^{4}(t) \equiv
\langle [ {\vec r}_{s}(t) - {\vec r}_{s}(0) ]^{4} \rangle$.
General properties of $\alpha_{2}(t)$ revealed by computer simulations
can be summarized as follows~\cite{Kob97}:
(i) on the time scale at which the motion of particles is
ballistic, $\alpha_{2}(t)$ is zero;
(ii) upon entering the time scale of the $\beta$ relaxation
where the density correlators are close to their plateaus, 
$\alpha_{2}(t)$ starts to increase;
and (iii) on the time scale of the $\alpha$ relaxation, 
$\alpha_{2}(t)$ decreases to zero, 
reflecting diffusive dynamics at long times which is a Gaussian process. 
It is also observed that 
the maximum value of $\alpha_{2}(t)$ and the time at which this maximum 
is attained both increase with decreasing $T$~\cite{Kob97}.
Positive $\alpha_{2}(t)$ means that the probability for a particle
to move very far is enhanced relative to the one expected for
a random-walk process.
The peak height of $\alpha_{2}(t)$ has therefore been interpreted
as a measure of the dynamical heterogeneity that reflects
different local environments around an individual particle.

\begin{figure}[tb]
\centerline{\includegraphics[width=0.8\linewidth]{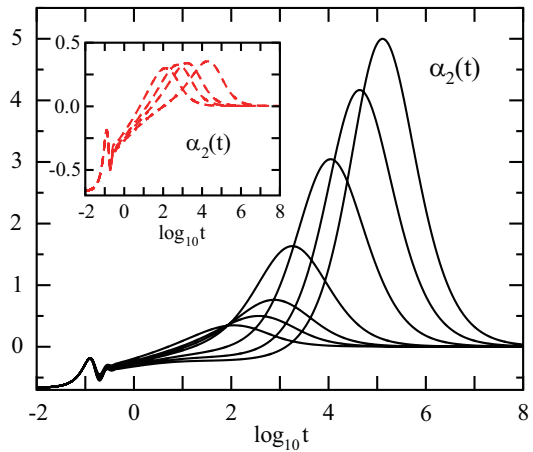}}
\caption{
(Color online)
Non-Gaussian parameter $\alpha_{2}(t)$ as a function of $\log_{10} t$
based on the extended MCT 
for reduced temperatures 
$\epsilon = (T_{\rm c} - T)/T_{\rm c} = -0.10$,
$-0.05$, $-0.03$, $-0.01$, $+0.01$, 
$+0.03$, and $+0.05$
(from left to right).
The corresponding results from the idealized MCT,
but for $\epsilon < 0$ (i.e., $T > T_{\rm c}$) only, are
presented as dashed curves 
in the inset to avoid the overcrowding of the figure.
Notice about an order of magnitude 
difference in the ordinate scales.}
\label{fig:NGP}
\end{figure}

The extended-MCT equations for determining the non-Gaussian parameter are
derived in Appendix~\ref{appendix:NGP}, and the resulting $\alpha_{2}(t)$
are plotted in Fig.~\ref{fig:NGP} 
for reduced temperatures $\epsilon = (T_{\rm c} - T)/T_{\rm c}$ 
specified in the caption.
The corresponding results from the idealized MCT,
but for $\epsilon < 0$ (i.e., $T > T_{\rm c}$) only, are
presented as dashed curves 
in the inset to avoid the overcrowding of the figure.
As noticed at the end of Appendix~\ref{appendix:NGP},
the peculiar behavior $\alpha_{2}(t \to 0) = -2/3$ for short times
as predicted by our theory 
simply reflects that the ideal-gas contribution to the memory kernel
is discarded in the mode-coupling approach.
Such an ideal-gas contribution is 
responsible for the short-time ballistic regime,  
but is irrelevant as far as the long-time dynamics is concerned.

It is seen from the inset of Fig.~\ref{fig:NGP} that
the peak height of $\alpha_{2}(t)$ predicted by the idealized MCT
does not grow with decreasing $T$, and it is underestimated
by almost an order of magnitude compared to that
reported in computer simulations~\cite{Kob97}.
This defect has already been known from the theoretical work in Ref.~\cite{Fuchs98},
and has been demonstrated more explicitly in Ref.~\cite{Flenner05b}.
The main panel of Fig.~\ref{fig:NGP}, on the other hand, indicates that
a substantial improvement is achieved by the extended MCT in that
the peak height grows upon lowering $T$ to an extent as 
observed in simulations~\cite{Kob97}. 
(Notice an order of magnitude difference in the ordinate scales
of the main panel and the inset.)
Thus, the extended MCT reproduces a feature of the dynamical
heterogeneity characterized by the non-Gaussian parameter,
and this is accomplished via the inclusion of the hopping processes.

\subsubsection{Probability distribution of particle displacements}

In terms of the non-Gaussian parameter $\alpha_{2}(t)$, 
the dynamics is most heterogeneous
in the late-$\beta$ or early $\alpha$ regime
where the density correlators start to decay from the plateau
(see the upper panel of Fig.~\ref{fig:phi}).
Recently, it has been recognized 
from studies of four-point density correlation functions that
there exists dynamical heterogeneity on a much longer time scale
comparable to the $\alpha$-relaxation time $\tau_{q^{*}}$~\cite{Lacevic03}.
Recognizing that $\alpha_{2}(t)$ is dominated by those particles
which move farther than expected from a Gaussian distribution 
of particle displacements, 
Flenner and Szamel introduced a new non-Gaussian parameter
$\gamma(t) \equiv 
(1/3) \delta r^{2}(t) [\delta r^{2}]^{-1}(t) - 1$
with 
$[\delta r^{2}]^{-1}(t) \equiv
\langle 1/ [ {\vec r}_{s}(t) - {\vec r}_{s}(0) ]^{2} \rangle$,
which weights strongly the particles which have not moved as 
far as expected from the Gaussian distribution~\cite{Flenner05}.
It is found that the peak position of $\gamma(t)$ is located 
on the time scale of $\tau_{q^{*}}$, and hence, characterizes
the longer-time dynamical heterogeneity.
It is also observed that the peak position of $\gamma(t)$ 
corresponds to the time at which 
two peaks in the probability distribution of particle 
displacements, reflecting populations of mobile and immobile particles (see below), 
are of equal height. 
This implies that the presence of the longer-time dynamical heterogeneity 
can be examined also through such a probability distribution. 

The mentioned probability distribution $P(\log_{10}(\delta r);t)$ of the
logarithm of particle displacements $\delta r$ at time $t$
can be obtained from the van Hove self correlation function
$G_{s}(\delta r,t)$ by the transformation~\cite{Flenner05}
\begin{equation}
P(\log_{10}(\delta r);t) =
\ln(10) \, 4 \pi \delta r^{3} G_{s}(\delta r,t).
\label{eq:Psr-def}
\end{equation}
The probability distribution is defined such that the
integral $\int_{x_{0}}^{x_{1}} dx \, P(x;t)$ is the fraction of particles
whose value of $\log_{10} (\delta r)$ is between $x_{0}$ and $x_{1}$. 
If the motion of a particle is diffusive with a diffusion coefficient $D$, 
there holds 
$G_{s}(\delta r,t) \approx
[ 1 / (4 \pi Dt)^{3/2} ] \exp( - \delta r^{2} / 4 D t)$~\cite{Hansen86}.
As argued in Ref.~\cite{Flenner05}, the 
shape of the corresponding probability distribution 
under the diffusion approximation becomes time independent
with the peak height $\approx 2.13$. 

\begin{figure}[tb]
\centerline{\includegraphics[width=0.8\linewidth]{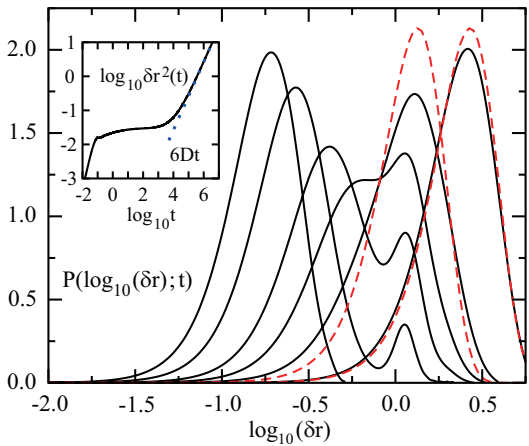}}
\caption{
Probability distribution $P(\log_{10} (\delta r);t)$ 
of the logarithm of single
particle displacements based on the extended MCT 
at the reduced temperature
$\epsilon = (T_{\rm c} - T)/T_{\rm c} = + 0.01$ (solid curves).
The times shown are 
$t = 5.4 \times 10^{3}$,
$4.3 \times 10^{4}$,
$1.7 \times 10^{5}$,
$3.4 \times 10^{5}$,
$6.8 \times 10^{5}$, and
$2.7 \times 10^{6}$
(from left to right).
The $\alpha$-relaxation time $\tau_{q^{*}}$ at this reduced temperature is
$\tau_{q^{*}} = 4.3 \times 10^{5}$ ({\em cf.} Fig.~\ref{fig:phi}). 
The dashed curves refer to the probability distribution under the
diffusion approximation for
$t = 6.8 \times 10^{5}$ and $2.7 \times 10^{6}$.
The inset exhibits the time evolution of 
the mean-squared displacement $\delta r^{2}(t)$ at $\epsilon = +0.01$
on double logarithmic scales (solid curve).
The dotted curve refers to the diffusion asymptote, $6 Dt$.}
\label{fig:Psr}
\end{figure}

Figure~\ref{fig:Psr} shows the extended-MCT result for 
the probability distribution $P(\log_{10} (\delta r);t)$ (solid curves)
at the reduced temperature $\epsilon = (T_{\rm c} - T)/T_{\rm c} = + 0.01$. 
The times shown are 
$t = 5.4 \times 10^{3}$,
$4.3 \times 10^{4}$,
$1.7 \times 10^{5}$,
$3.4 \times 10^{5}$,
$6.8 \times 10^{5}$, and
$2.7 \times 10^{6}$.
The $\alpha$-relaxation time $\tau_{q^{*}}$ at this reduced temperature is
$\tau_{q^{*}} = 4.3 \times 10^{5}$ ({\em cf.} Fig.~\ref{fig:phi}). 
The dashed curves refer to the probability distribution under the
diffusion approximation for
$t = 6.8 \times 10^{5}$ and $2.7 \times 10^{6}$.
The inset exhibits the mean-squared displacement $\delta r^{2}(t)$ at 
$\epsilon = +0.01$ on double logarithmic scales,
from which one understands that the times chosen 
in the main panel range from the late-$\beta$, plateau regime to the
final $\alpha$ regime where $\delta r^{2}(t) \sim 6Dt$.

The appearance of the plateau in $\delta r^{2}(t)$ is due to particles
being caged~\cite{Fuchs98}, 
and the peak of the probability distribution $P(\log_{10} (\delta r);t)$ 
for $t = 5.4 \times 10^{3}$ reflects populations of such ``immobile'' particles.
At later times, the second peak develops in the probability distribution
at $\log_{10} (\delta r) \approx 0.0$ (i.e., $\delta r \approx 1.0$),
reflecting ``mobile'' particles hopping over interparticle distances.
One infers from Fig.~\ref{fig:Psr} that 
the time at which the two peaks in $P(\log_{10} (\delta r);t)$
become of equal height is located on the time scale of $\tau_{q^{*}}$, and this
is consistent with the simulation result~\cite{Flenner05}.
Subsequently,  the double-peak structure disappears, and 
the probability distribution approaches the one well described
by the diffusion approximation.

Coexistence of mobile and immobile particles 
is a direct indication of the dynamical heterogeneity~\cite{Ediger00}.
Our results shown in Fig.~\ref{fig:Psr} indicate that 
the hopping processes are responsible for such a double-peak
structure in the probability distribution
$P(\log_{10} (\delta r);t)$ 
occurring on the time scale $\tau_{q^{*}}$ of the $\alpha$ relaxation. 
This also explains why the idealized MCT failed to reproduce the double-peak
structure in $P(\log_{10} (\delta r);t)$~\cite{Flenner05b}.

\subsubsection{Growing dynamic length scale}
\label{subsec:length}

Figure~\ref{fig:Psr} also implies that the probability distribution
approaches its diffusion asymptote 
only after its peak position exceeds those length scales 
where the double-peak structure in $P(\log_{10} (\delta r);t)$ 
is observable.
Thus, there is a certain length scale above which Fickian diffusion sets in.
In the following, we shall quantify such a length scale characterizing the
crossover from non-Fickian to Fickian diffusion, and investigate
its temperature dependence.

To this end, let us introduce the ratio $R_{q}^{s}$ of the
product $q^{2} D \tau_{q}^{s}$ to the one at some reference
temperature, 
which is an analogue of the ratio $R$ studied in the inset of
Fig.~\ref{fig:Dtau}.
Here the $\alpha$-relaxation time $\tau_{q}^{s}$
of the tagged-particle density correlator is defined via the convention
$\phi_{q}^{s}(\tau_{q}^{s}) = 0.1$.
Under the diffusion approximation, there holds
$\phi_{q}^{s}(t) \approx \exp( - q^{2} D t )$~\cite{Hansen86}.
Thus, the product $q^{2} D \tau_{q}^{s}$ 
is constant, and hence, the ratio $R_{q}^{s}$ is unity,
if the dynamics on the length scale 
$\approx 2\pi / q$ is diffusive.
On the other hand, the ratio $R_{q}^{s}$ exceeds unity if
the dynamics on the length scale $\approx 2\pi / q$ 
is non-Fickian.
We shall therefore define the crossover wave number $q_{\rm onset}$
via $R_{q_{\rm onset}}^{s} = 1.1$, i.e.,
as the wave number at which the ratio $R_{q}^{s}$ reaches 
10\% above unity. 
The onset length scale of Fickian diffusion shall then be defined
via $\ell_{\rm onset} \equiv 2 \pi / q_{\rm onset}$. 

\begin{figure}[tb]
\centerline{\includegraphics[width=0.9\linewidth]{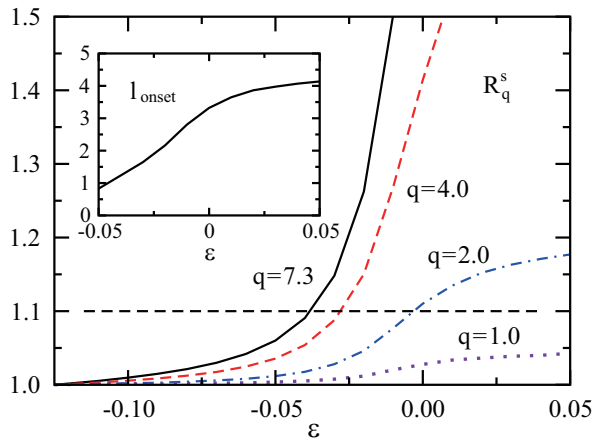}}
\caption{
(Color online)
Ratio $R_{q}^{s}$ of the product $q^{2} D \tau_{q}^{s}$ to the one at 
a reference temperature $T_{\rm ref}$ 
as a function of the reduced temperature 
$\epsilon = (T_{\rm c}-T)/T_{\rm c}$.
Here $T_{\rm ref}$ is chosen such that 
$(T_{\rm c}-T_{\rm ref})/T_{\rm c} = -0.125$.
The wave numbers shown are $q = 7.3$ (solid curve),
4.0 (dashed curve), 2.0 (dashed-dotted curve), and
1.0 (dotted curve).
The horizontal dashed line marks $R_{q}^{s} = 1.1$
chosen to determine the crossover wave number $q_{\rm onset}$ 
introduced in the text.
The inset exhibits the onset length scale of Fickian diffusion 
$\ell_{\rm onset} \equiv 2 \pi / q_{\rm onset}$ as a function of the
reduced temperature $\epsilon$.}
\label{fig:length}
\end{figure}

Figure~\ref{fig:length} exhibits the ratio $R_{q}^{s}$
for wave numbers $q = 7.3$ (solid curve), 4.0 (dashed curve),
2.0 (dashed-dotted curve), and
1.0 (dotted curve), 
as a function of the reduced temperature 
$\epsilon = (T_{\rm c}-T)/T_{\rm c}$.
The reference temperature $T_{\rm ref}$ for calculating $R_{q}^{s}$ 
is chosen such that 
$(T_{\rm c}-T_{\rm ref})/T_{\rm c} = -0.125$.
The horizontal dashed line marks $R_{q}^{s} = 1.1$ 
chosen to determine the crossover wave number $q_{\rm onset}$
introduced above. 

For reduced temperatures $\epsilon \lesssim -0.04$,
all the ratios $R_{q}^{s}$ shown in Fig.~\ref{fig:length} are less than 1.1,
implying that the tagged-particle dynamics on the length scales comparable to 
and larger than
$2\pi / 7.3 \approx 0.86$ can be well described by the Fickian diffusion law.
For $\epsilon > -0.04$, $R_{q}^{s}$ for $q = 7.3$ exceeds 1.1,
meaning that the dynamics is diffusive only on length scales larger than 0.86.
The onset length scale $\ell_{\rm onset}$ of Fickian diffusion increases to 
$2\pi/4.0 \approx 1.6$ at $\epsilon \approx -0.03$, and to
$2\pi/2.0 \approx 3.1$ at $\epsilon \approx 0$, 
as can be inferred from Fig.~\ref{fig:length}.
The length scale $\ell_{\rm onset}$ so obtained as a function of the
reduced temperature $\epsilon$ is summarized in the inset of Fig.~\ref{fig:length}.
Thus, the extended MCT predicts the presence of a growing dynamic length scale.
It is anticipated that $\ell_{\rm onset}$ is intimately related to the
mean size of the dynamic clusters observed in simulations~\cite{Berthier04},
since Fickian diffusion, i.e., a random-walk process,
is possible only over those length scales where 
the coherence length of such clusters is smeared out.

\section{Summary}
\label{sec:conclusion}

In this paper, we developed a new extended version of MCT for glass transition.
The activated hopping processes are incorporated via the dynamical theory,
originally formulated to describe diffusion-jump processes in crystals
and adapted in this work to glass-forming liquids.
The dynamical-theory approach treats hopping as arising from 
vibrational fluctuations in quasi-arrested state where particles are trapped
inside their cages, and the hopping rate is formulated
in terms of the Debye-Waller factors characterizing
the structure of the quasi-arrested state. 
The resulting expression for the hopping rate
takes an activated form, and 
the barrier height for the hopping is ``self-generated'' in the sense 
that it is present only in those states where
the dynamics exhibits a well defined plateau.
It is discussed how such a hopping rate can be incorporated to 
develop a new extended MCT.

The extended MCT deals with the interplay of two effects.
Nonlinear interactions of density fluctuations, as described by the idealized
memory kernel $m_{q}^{\rm id}(z)$, lead to the cage effect with a trend to
produce arrested states for sufficiently low temperatures. 
Phonon assisted hoppings, taken into account via the hopping kernel
$\delta_{q}(z)$, lead to the $\alpha$ relaxation at all temperatures and restore ergodicity.
The interplay of these two effects is described by the memory kernel $m_{q}(z)$
via Eq.~(\ref{eq:extended-memory-b})
which has the form of a Dyson equation.
As demonstrated in Sec.~\ref{sec:results}
for a model of the Lennard-Jones system,
it leads to nontrivial theoretical predictions concerning the breakdown of the
SE relation and characteristic features of dynamical heterogeneities, 
which the idealized version of the theory failed to reproduce.

In dense liquids, relaxation is necessarily connected with rearrangements of
large complexes of particles.
We have seen in Fig.~\ref{fig:NGP} that the peak height of the
non-Gaussian parameter $\alpha_{2}(t)$ grows substantially
in the late-$\beta$ regime as the temperature is decreased.
This enhanced probability for a particle to move further
is what one would expect as a result of the
building of a backflow in the liquid, and
it is anticipated that the string-like motions observed in the 
late-$\beta$ regime~\cite{Donati98}
reflect such a backflow pattern.
The backflow was originally discussed by Feynman and Cohen
and taken into account in their theory
for roton spectrum in liquid helium~\cite{Feynman56}.
Subsequently, 
it was found that the quantum-mechanical analogue of the idealized memory
kernel reproduces the Feynman-Cohen result
for roton spectrum~\cite{Goetze76}, i.e.,
the backflow phenomenon 
is within the scope of the idealized MCT~\cite{Goetze95}. 
However, the idealized memory kernel alone does not account for such
a pronounced peak in $\alpha_{2}(t)$ as observed in computer 
simulations~\cite{Fuchs98}.
This implies that the mentioned interplay with the 
hopping kernel plays a relevant role for the substantial increase of $\alpha_{2}(t)$
and the building of the backflow. 

The double-peak structure in the probability distribution $P(\log_{10} (\delta r);t)$ 
of particle displacements, which is most pronounced on the 
time scale $\tau_{q^{*}}$ of the $\alpha$-relaxation and disappears 
at longer times ({\em cf.} Fig.~\ref{fig:Psr}),
reflects coexistence of mobile and immobile particles with a life time $\approx \tau_{q^{*}}$, 
and is a direct indication of the dynamical heterogeneity~\cite{Ediger00}.
The extended MCT developed here provides a natural explanation for its origin
in terms of the cage and hopping effects.
As argued in Sec.~\ref{subsec:length}, the life time of the double-peak structure
gives rise to a growing dynamic length scale $\ell_{\rm onset}$,
which is associated with the decoupling of the time scales of the $\alpha$ processes
occurring on length scales smaller than $\ell_{\rm onset}$ from the
diffusion coefficient.
The breakdown of the SE relation discussed in Sec.~\ref{subsec:breakdown}
can also be understood 
in terms of the decoupling of the $\alpha$-relaxation
time $\tau_{q^{*}}$ from the diffusivity
which occurs when $\ell_{\rm onset}$ exceeds the 
average interparticle distance $\approx 2\pi/q^{*}$. 
It is anticipated that the onset length scale $\ell_{\rm onset}$ of Fickian diffusion
is intimately related to the
mean size of the dynamic clusters observed in simulations~\cite{Berthier04},
since Fickian diffusion, i.e., a random-walk process,
is possible only over those length scales where 
the coherence length of such clusters is smeared out.
Then, our picture for the breakdown of the SE relation is consistent with
that of Ref.~\cite{Berthier04}, where the connection between decoupling
phenomena and the growing coherence length scale is discussed.

It is certainly necessary to improve further the theory developed here.
First of all, possible correlated hopping effects are discarded in our model
for the hopping kernel.
This might explain why our theory underestimates the degree of the breakdown
of the SE relation near $T_{\rm g}$, which we referred 
to in connection with Fig.~\ref{fig:Dtau}.
That the development of the length scale $\ell_{\rm onset}$
seems suppressed for $\epsilon > 0$, as can be inferred from 
the inset of Fig.~\ref{fig:length}, 
might also be related to such a defect of the present theory.
Second, we did not determine $\Delta^{2}$ --
a square of the ratio $\Delta = x^{*}/a$ formed with the critical 
size $x^{*}$ of the phonon-assisted fluctuation needed to cause a hopping 
and the average interparticle distance $a$ ({\em cf.} Sec.~\ref{sec:theory}) --
microscopically, but simply took a value ($\Delta^{2} = 0.10$) from the literature.
Regarding this, let us mention that 
nearly the same results as those presented in Sec.~\ref{sec:results}
can be obtained with other values of $\Delta^{2}$,
as far as those values 
consistent with the Lindemann length~\cite{comment-Lindemann-length} are chosen.
Thus, our theoretical results do not rely on the specific value of $\Delta^{2} = 0.10$.
But, of course, it is desirable to determine $\Delta^{2}$ consistently within the theory.
Third, the $\alpha$-relaxation stretching is not enhanced at low temperatures.
For example, the stretching exponents $\beta_{q}$, obtained via 
Kohlrausch-law fits of the density correlators shown in Fig.~\ref{fig:phi},
for $q = 10.0$ are
0.76, 0.79, and 0.80 for $\epsilon = +0.01$, $+0.05$, and $+0.10$, respectively.
One possibility to overcome this defect is to take into account
distribution of hopping rates or barrier heights, which can be done, 
e.g., by considering fluctuations in the Debye-Waller factors and $\Delta^{2}$.
Four-point density correlators might be necessary for this purpose, 
though it is a difficult task to calculate such higher-order correlators.
But, in view of the significant results achieved by our theory,
it is promising to pursue its further development.

\begin{acknowledgments}

The author is grateful to W.~G\"otze and B.~Kim for discussions.
He also thanks H.~Sillescu for sending him the
diffusivity data of salol presented in Ref.~\cite{Chang97},
and W.~Kob for the simulation results of Ref.~\cite{Kob94}. 
This work was supported 
by Grant-in-Aids for scientific 
research from the 
Ministry of Education, Culture, Sports, Science and 
Technology of Japan (No.~20740245).

\end{acknowledgments}

\appendix

\section{Novel derivation of the hopping kernel of G\"otze and Sj\"ogren}
\label{appendix:GS-theory}

In order to go beyond the idealized MCT, one has to consider
corrections to the idealized memory kernel,
$m_{q}(z) = m_{q}^{\rm id}(z) + \Delta m_{q}(z)$.
The extended MCT of G\"otze and Sj\"ogren~\cite{Goetze87,Goetze95}
can be considered as a theory for such corrections, and 
yields an expression for $\Delta m_{q}(z)$ in terms of the 
hopping kernel $\delta_{q}(z)$ that involves couplings to currents. 
Their extended theory is formulated with the generalized kinetic
theory for phase-space density fluctuations, but essentially the same 
expression for the hopping kernel $\delta_{q}(z)$ can be
derived also with the standard projection-operator approach
based on density and current-density fluctuations.
In this appendix, we outline such a novel derivation.

We start from reviewing the derivation of the idealized memory
kernel $m_{q}^{\rm id}(t)$ (see Ref.~\cite{Goetze91b} for details). 
Let us introduce a projection operator ${\cal P}$ onto 
the subspace spanned by the density fluctuations $\rho_{\vec q}$
and the current-density fluctuations
$j_{\vec q}^{\lambda} \equiv \sum_{i} v_{i}^{\lambda} \exp( i {\vec q} \cdot {\vec r}_{i} )$
($\lambda = x, y, z$).
Here $v_{i}^{\lambda}$ denotes the $\lambda$ component of the velocity of $i$th particle.
Within the Zwanzig-Mori formalism~\cite{Hansen86},
one obtains based on the operator ${\cal P}$ 
the exact equation (\ref{eq:GLE-phi-a}) for the coherent density correlator $\phi_{q}(t)$.
The formal expression for the memory kernel entering there reads
\begin{equation}
\Omega_{q}^{2} m_{q}(t) = 
\frac{M}{N k_{\rm B}T}
\langle F_{\vec q}^{*} e^{i {\cal QLQ} t} F_{\vec q} \rangle,
\end{equation}
where ${\cal Q} \equiv 1 - {\cal P}$, and the fluctuating force is given by
\begin{equation}
F_{\vec q} = \partial_{t} (\hat{\vec q} \cdot {\vec j}_{\vec q})
- iq \frac{k_{\rm B}T}{M S_{q}} \rho_{\vec q}.
\end{equation}
Here $\hat{\vec q} = {\vec q} / q$. 
Under the mode-coupling approach, the fluctuating forces are approximated
by their projections onto the subspace spanned by pair-density modes
$\rho_{\vec k} \rho_{\vec p}$;
this is done by introducing the second projection operator 
\begin{equation}
{\cal P}_{2} X =
\sum_{{\vec k} > {\vec p}}
\rho_{\vec k} \rho_{\vec p} \,
\frac{1}{N^{2} S_{k} S_{p}} \,
\langle \rho_{\vec k}^{*} \rho_{\vec p}^{*} X \rangle,
\label{eq:P2-def}
\end{equation}
and approximating $F_{\vec q} \approx {\cal P}_{2} F_{\vec q} \equiv F_{\vec q}^{\rho \rho}$, 
i.e., 
\begin{equation}
\Omega_{q}^{2} m_{q}(t) \approx
\frac{M}{N k_{\rm B}T}
\langle F_{\vec q}^{\rho \rho \, *} e^{i {\cal QLQ} t} F_{\vec q}^{\rho \rho} \rangle.
\label{eq:m-before-factorization}
\end{equation}
Notice here that the {\em static} factorization approximation
\begin{equation}
\langle \rho_{\vec k}^{*} \rho_{\vec p}^{*} 
\rho_{{\vec k}^{\prime}} \rho_{{\vec p}^{\prime}} \rangle \approx
\delta_{{\vec k}, {\vec k}^{\prime}}
\delta_{{\vec p}, {\vec p}^{\prime}}
N^{2} S_{k} S_{p} 
\,\,\,
({\vec k} > {\vec p}, {\vec k}^{\prime} > {\vec p}^{\, \prime}),
\label{eq:static-factorization}
\end{equation}
 is already used in the definition (\ref{eq:P2-def}) of ${\cal P}_{2}$.
Within the convolution approximation for triple density correlations, one finds
for the projected fluctuating force
\begin{subequations}
\begin{eqnarray}
F_{\vec q}^{\rho \rho} &=&
- i \frac{\rho k_{\rm B}T}{NM}
\sum_{{\vec k} > {\vec p}}
\delta_{{\vec q}, {\vec k}+{\vec p}} 
\nonumber \\
& &
\qquad \qquad 
\times \, 
( \hat{\vec q} \cdot {\vec k} \, c_{k} + \hat{\vec q} \cdot {\vec p} \, c_{p} ) 
\rho_{\vec k} \rho_{\vec p}
\label{eq:projected-F-a}
\\
&=&
- i \frac{\rho k_{\rm B}T}{NM}
\sum_{\vec k}
\hat{\vec q} \cdot {\vec k} \, c_{k} \rho_{\vec k} \rho_{{\vec q}-{\vec k}}.
\label{eq:projected-F-b}
\end{eqnarray}
\end{subequations}
Substituting Eq.~(\ref{eq:projected-F-a}) into Eq.~(\ref{eq:m-before-factorization}) and then
using the {\em dynamical} factorization approximation
\begin{eqnarray}
& &
\langle \rho_{\vec k}^{*} \rho_{\vec p}^{*} \, e^{i {\cal QLQ} t} 
\rho_{{\vec k}^{\prime}} \rho_{{\vec p}^{\prime}} \rangle \approx
\delta_{{\vec k}, {\vec k}^{\prime}}
\delta_{{\vec p}, {\vec p}^{\prime}}
N^{2} S_{k} S_{p} \phi_{k}(t) \phi_{p}(t)
\nonumber \\
& & 
\qquad \qquad
\qquad \qquad
\qquad \qquad
({\vec k} > {\vec p}, {\vec k}^{\prime} > {\vec p}^{\, \prime}),
\label{eq:dynamic-factorization}
\end{eqnarray}
which factorizes averages of products evolving in time with the
generator ${\cal QLQ}$ into products of averages formed
with variables evolving with ${\cal L}$, 
one obtains the idealized kernel $m_{q}^{\rm id}(t)$
given in Eqs.~(\ref{eq:ideal-MCT-m}).

To extend the idealized theory, one has to avoid the use of the
dynamical factorization approximation~(\ref{eq:dynamic-factorization}).
We shall therefore start from 
Eqs.~(\ref{eq:m-before-factorization}) and (\ref{eq:projected-F-b})
for the memory kernel $m_{q}(t)$, to which 
the approximation~(\ref{eq:dynamic-factorization}) has yet to be applied.
Notice, however, that the static factorization approximation~(\ref{eq:static-factorization})
is still assumed in our approach; this is employed in the definition~(\ref{eq:P2-def})
of ${\cal P}_{2}$, which is then used in deriving Eq.~(\ref{eq:projected-F-b}).
Thus, there holds $m_{q}(0) = m_{q}^{\rm id}(0)$ at $t = 0$.

By introducing a new projection operator ${\cal P}^{\prime}$ onto the 
subspace spanned by $F_{\vec q}^{\rho \rho}$, i.e.,
\begin{equation}
{\cal P}^{\prime} X =
F_{\vec q}^{\rho \rho} 
\frac{M}{N k_{\rm B}T \, \Omega_{q}^{2} m_{q}(0)}
\langle
F_{\vec q}^{\rho \rho \, *} X \rangle,
\end{equation}
one obtains the following Zwanzig-Mori equation of motion for $m_{q}(t)$:
\begin{equation}
\partial_{t} m_{q}(t) + m_{q}(0) \int_{0}^{t} dt^{\prime} \,
L_{q}(t-t^{\prime}) m_{q}(t^{\prime}) = 0.
\label{eq:GLE-m}
\end{equation}
Here the factor $m_{q}(0)$ in front of the convolution integral is just for later convenience,
and the formal expression for the ``memory kernel'' $L_{q}(t)$ is given by
\begin{equation}
m_{q}(0) L_{q}(t) = 
\frac{M}{N k_{\rm B}T \, \Omega_{q}^{2} m_{q}(0)}
\langle R_{\vec q}^{*} \, e^{i {\cal Q}^{\prime} {\cal QLQ} {\cal Q}^{\prime} t} \,
R_{\vec q} \rangle,
\label{eq:L-def}
\end{equation}
in terms of the ``fluctuating force''
\begin{equation}
R_{\vec q} = i {\cal Q}^{\prime} {\cal QLQ} F_{\vec q}^{\rho \rho},
\label{eq:R-def}
\end{equation}
where we have introduced ${\cal Q}^{\prime} \equiv 1 - {\cal P}^{\prime}$.
Substituting Eq.~(\ref{eq:projected-F-b}) into Eq.~(\ref{eq:R-def}),
one finds
\begin{eqnarray}
R_{\vec q} &=&
\frac{\rho k_{\rm B}T}{NM}
\sum_{\vec k}
\hat{\vec q} \cdot {\vec k} \, c_{k}
\bigl\{ 
  {\vec k} \cdot {\vec j}_{\vec k} \, \rho_{\vec p} 
  + {\vec p} \cdot {\vec j}_{\vec p} \, \rho_{\vec k}
\nonumber \\
& & 
\qquad \qquad \quad 
  - \,
  ( {\vec k} \cdot {\vec j}_{\vec q} \, S_{p}
  + {\vec p} \cdot {\vec j}_{\vec q} \, S_{k})
\bigr\},
\end{eqnarray}
in which ${\vec p} = {\vec q} - {\vec k}$.

Now, we apply the mode-coupling approximation to the kernel $L_{q}(t)$.
Because of the odd time reversal symmetry of $R_{\vec q}$,
the simplest mode-coupling approximation for $L_{q}(t)$ can be
introduced by defining the projection operator ${\cal P}_{2}^{\prime}$ 
onto the subspace panned by the product
$\rho_{\vec k} j_{\vec p}^{\lambda}$
of the density and current-density modes:
\begin{equation}
{\cal P}_{2}^{\prime} X =
\sum_{\vec k} \sum_{\lambda}
\rho_{\vec k} j_{\vec p}^{\lambda} \, 
\frac{M}{N^{2} k_{\rm B}T \, S_{k}} \,
\langle \rho_{\vec k}^{*} j_{\vec p}^{\lambda \, *} X \rangle,
\label{eq:P2-prime-def}
\end{equation}
where we have used the factorization approximation
\begin{equation}
\langle \rho_{\vec k}^{*} j_{\vec p}^{\lambda \, *} 
\rho_{{\vec k}^{\prime}} j_{{\vec p}^{\prime}}^{\mu} \rangle \approx
\delta_{{\vec k}, {\vec k}^{\prime}}
\delta_{\lambda \mu}
N^{2} k_{\rm B}T S_{k} / M.
\end{equation}
Here and in the following, we use abbreviations
${\vec p} \equiv {\vec q} - {\vec k}$ and 
${\vec p}^{\, \prime} \equiv {\vec q} - {\vec k}^{\prime}$.
We thus obtain under the mode-coupling approximation
$R_{\vec q} \approx {\cal P}_{2}^{\prime} R_{\vec q} \equiv R_{\vec q}^{\rho j}$, 
\begin{equation}
m_{q}(0) L_{q}(t) \approx
\frac{M}{N k_{\rm B}T \, \Omega_{q}^{2} m_{q}(0)}
\langle R_{\vec q}^{\rho j \, *} 
e^{i {\cal Q}^{\prime} {\cal QLQ} {\cal Q}^{\prime} t}
R_{\vec q}^{\rho j} \rangle.
\label{eq:MCT-L-0}
\end{equation}
Within the convolution approximation for triple density correlations, one finds
\begin{subequations}
\begin{equation}
R_{\vec q}^{\rho j} =
\frac{\rho k_{\rm B}T}{NM}
\sum_{\vec k} \sum_{\lambda}
V_{\lambda}({\vec q}; {\vec k}, {\vec p} \, ) \, 
\rho_{\vec k} j_{\vec p}^{\lambda},
\label{eq:MCT-R-a}
\end{equation}
in which $V_{\lambda}$ is given by
\begin{eqnarray}
& &
V_{\lambda}({\vec q}; {\vec k}, {\vec p} \, ) =
\frac{1}{(2 \pi)^{3}} \int d{\vec k}^{\prime} \,
\hat{\vec q} \cdot {\vec k}^{\prime} \, c_{k^{\prime}} 
\nonumber \\
& & 
\qquad \quad
\times \, 
( k_{\lambda}^{\prime} S_{p^{\prime}}
  h_{|{\vec p}^{\prime}-{\vec k}|} +
  p_{\lambda}^{\prime} S_{k^{\prime}}
  h_{|{\vec k}^{\prime}-{\vec k}|} ).
\label{eq:MCT-R-b}
\end{eqnarray}
\end{subequations}
Here $k_{\lambda}$ refers to the $\lambda$ component of the vector
${\vec k}$, and the 
function $h_{q}$ is given by $h_{q} = c_{q} S_{q}$.
When Eq.~(\ref{eq:MCT-R-a}) is substituted into Eq.~(\ref{eq:MCT-L-0}),
the kernel $L_{q}(t)$ is expressed in terms of 
four-mode correlators, for which we invoke the following dynamical
factorization approximation:
\begin{eqnarray}
& &
\langle \rho_{\vec k}^{*} j_{\vec p}^{\lambda \, *}
e^{i {\cal Q}^{\prime} {\cal QLQ} {\cal Q}^{\prime} t}
\rho_{{\vec k}^{\prime}} j_{{\vec p}^{\prime}}^{\mu} \rangle
\nonumber \\
& &
\approx
\delta_{{\vec k}, {\vec k}^{\prime}}
\langle \rho_{\vec k}^{*} e^{i {\cal L} t} \rho_{\vec k} \rangle 
\langle j_{\vec p}^{\lambda \, *} e^{i {\cal L} t} j_{\vec p}^{\mu} \rangle 
\nonumber \\
& &
\quad
+ \,
\delta_{{\vec k}, {\vec p}^{\prime}}
\langle \rho_{\vec k}^{*} e^{i {\cal L} t} j_{\vec k}^{\mu} \rangle
\langle  j_{\vec p}^{\lambda \, *} e^{i {\cal L} t} \rho_{\vec p} \rangle
\nonumber \\
& &
=
\delta_{{\vec k}, {\vec k}^{\prime}}
N^{2} (k_{\rm B}T / M) S_{k} \phi_{k}(t) 
\nonumber \\
& &
\quad \qquad
\times \, 
[ \hat{p}_{\lambda} \hat{p}_{\mu} \phi_{p}^{\rm L}(t) +
  (\delta_{\lambda \mu} - \hat{p}_{\lambda} \hat{p}_{\mu}) \phi_{p}^{\rm T}(t) ]
\nonumber \\
& &
\quad
- \,
\delta_{{\vec k}, {\vec p}^{\prime}}
N^{2} S_{k} S_{p} 
(\hat{k}_{\mu} \hat{p}_{\lambda} /kp) 
\partial_{t} \phi_{k}(t) \partial_{t} \phi_{p}(t).
\end{eqnarray} 
Here $\phi_{q}^{\rm L}(t)$ and $\phi_{q}^{\rm T}(t)$ are
longitudinal and transversal current correlators, respectively,
which are normalized to unity at $t = 0$.
We then obtain the following form for the kernel $L_{q}(t)$
that involves couplings to current modes: 
\begin{subequations}
\label{eq:MCT-L}
\begin{eqnarray}
L_{q}(t) &=& \int d{\vec k} \,
\bigl[
  V_{\rm L}({\vec q}; {\vec k}, {\vec p} \, ) \phi_{k}(t) \phi_{p}^{\rm L}(t) 
\nonumber \\
& &
\quad
+ \,
  V_{\rm T}({\vec q}; {\vec k}, {\vec p} \, ) \phi_{k}(t) \phi_{p}^{\rm T}(t) 
\nonumber \\
& &
\quad
+ \,
  V^{\prime}({\vec q}; {\vec k}, {\vec p} \, ) 
  \partial_{t} \phi_{k}(t) \, \partial_{t} \phi_{p}(t)
\bigr]. 
\label{eq:MCT-L-a}
\end{eqnarray}
The vertex functions $V_{\rm L}$, $V_{\rm T}$, and $V^{\prime}$ are expressed 
in terms of the thermal velocity, the static equilibrium quantities,
and $V_{\lambda}$ given in Eq.~(\ref{eq:MCT-R-b}) as
\begin{equation}
V_{\rm L}({\vec q}; {\vec k}, {\vec p} \, ) = 
\frac{\rho k_{\rm B}T}{(2 \pi)^{3} M} \frac{S_{q} S_{k}}{q^{2} m_{q}(0)^{2}}
\Bigl[ \sum_{\lambda} \hat{p}_{\lambda} V_{\lambda}({\vec q}; {\vec k}, {\vec p} \, ) \Bigr]^{2},
\end{equation}
\begin{eqnarray}
V_{\rm T}({\vec q}; {\vec k}, {\vec p} \, ) &=&
\frac{\rho k_{\rm B}T}{(2 \pi)^{3} M} \frac{S_{q} S_{k}}{q^{2} m_{q}(0)^{2}}
\Bigl\{
  \sum_{\lambda} V_{\lambda}({\vec q}; {\vec k}, {\vec p} \, )^{2} 
\nonumber \\
& &
\qquad
  - \,
  \Bigl[ \sum_{\lambda} \hat{p}_{\lambda} V_{\lambda}({\vec q}; {\vec k}, {\vec p} \, ) \Bigr]^{2}
\Bigr\},
\end{eqnarray}
\begin{eqnarray}
V^{\prime}({\vec q}; {\vec k}, {\vec p} \, ) &=& 
- \frac{\rho}{(2 \pi)^{3}} \frac{S_{q} S_{k} S_{p}}{q^{2} m_{q}(0)^{2}}
\Bigl[ \sum_{\lambda} \frac{\hat{k}_{\lambda}}{k} 
V_{\lambda}({\vec q}; {\vec p}, {\vec k} \, ) \Bigr] 
\nonumber \\
& &
\qquad
\times \,
\Bigl[ \sum_{\lambda} \frac{\hat{p}_{\lambda}}{p} 
V_{\lambda}({\vec q}; {\vec k}, {\vec p} \, ) \Bigr].
\end{eqnarray}
\end{subequations}

Finally, let us connect the Laplace transform of $L_{q}(t)$ 
to the hopping kernel $\delta_{q}(z)$ using the
definition given by Eq.~(\ref{eq:extended-memory-a}).
The Laplace transform of Eq.~(\ref{eq:GLE-m}) reads
\begin{equation}
m_{q}(z) = - \frac{m_{q}(0)}{z + m_{q}(0) L_{q}(z)}.
\label{eq:Laplace-GLE-m}
\end{equation}
In view of this expression, 
let us formally introduce the function $L_{q}^{\rm id}(z)$ in terms
of the Laplace transform of the idealized memory kernel $m_{q}^{\rm id}(z)$ via
\begin{equation}
m_{q}^{\rm id}(z) = - \frac{m_{q}(0)}{z + m_{q}(0) L_{q}^{\rm id}(z)}.
\label{eq:Laplace-GLE-m-ideal}
\end{equation}
Here we have used the equality $m_{q}(0) = m_{q}^{\rm id}(0)$ at $t=0$
noticed above.
Substituting Eqs.~(\ref{eq:Laplace-GLE-m}) and (\ref{eq:Laplace-GLE-m-ideal}) 
into Eq.~(\ref{eq:extended-memory-a}), one obtains for the hopping kernel
\begin{equation}
\delta_{q}(z) = L_{q}(z) - L_{q}^{\rm id}(z).
\label{eq:delta-appendix}
\end{equation}
From the functional form of $L_{q}(t)$ given in Eq.~(\ref{eq:MCT-L-a}) and interpreting
that $L_{q}^{\rm id}(z)$ in Eq.~(\ref{eq:delta-appendix}) subtracts those contributions
already accounted for by the idealized memory kernel
$m_{q}^{\rm id}(z)$, one understands that
the expression~(\ref{eq:delta-appendix}) for the hopping kernel 
is essentially the same as the one derived by
G\"otze and Sj\"ogren~\cite{Goetze87,Goetze95}.

\section{Extended MCT equations for
the mean-squared displacement and the non-Gaussian parameter}
\label{appendix:MSD-NGP}

\subsection{Mean-squared displacement}
\label{appendix:MSD}

The equation of motion for the mean-squared displacement
$\delta r^{2}(t)$ can be obtained
from Eq.~(\ref{eq:GLE-phi-a}) for $\phi_{q}^{s}(t)$ 
by exploiting its relation to the
small-$q$ behavior of
$\phi_{q}^{s}(t) = 1 - q^{2} \delta r^{2}(t) / 6 + O(q^{4})$~\cite{Hansen86}:
\begin{equation}
\partial_{t} \delta r^{2}(t) + 
\frac{k_{\rm B}T}{M} 
\int_{0}^{t} dt^{\prime} \, m_{(0)}^{s}(t-t^{\prime}) \, \delta r^{2}(t^{\prime}) = 
6 \frac{k_{\rm B}T}{M} t.
\label{eq:GLE-MSD}
\end{equation}
Here we have introduced the $q \to 0$ limit of the memory kernel via
$m_{(0)}^{s}(t) \equiv \lim_{q \to 0} q^{2} m_{q}^{s}(t)$.
To obtain $m_{(0)}^{s}(t)$, it is more convenient to rewrite 
Eq.~(\ref{eq:extended-memory-b}) in the form
\begin{equation}
m_{q}^{s}(z) =  m_{q}^{s \, {\rm id}}(z) +
m_{q}^{s \, {\rm id}}(z) \, \delta_{q}^{s}(z) \, m_{q}^{s}(z),
\label{eq:extended-memory-for-appendix}
\end{equation}
from which one finds 
\begin{subequations}
\label{eq:MCT-MSD}
\begin{equation}
m_{(0)}^{s}(z) = m_{(0)}^{s \, {\rm id}}(z) + 
m_{(0)}^{s \, {\rm id}}(z) \, \delta_{(0)}^{s}(z) \, m_{(0)}^{s}(z).
\label{eq:MCT-MSD-a}
\end{equation}
Here $m_{(0)}^{s \, {\rm id}}(z)$ refers to the corresponding $q \to 0$ limit of the
idealized memory kernel, and $\delta_{(0)}^{s}(z)$ is defined via
$\delta_{(0)}^{s}(z) \equiv \lim_{q \to 0} \delta_{q}^{s}(z) / q^{2}$.
The former is given by~\cite{Fuchs98}
\begin{equation}
m_{(0)}^{s \, {\rm id}}(t) = \frac{1}{6 \pi^{2}}
\int dk \, k^{4} \rho S_{k} c_{k}^{2}
\phi_{k}(t) \phi_{k}^{s}(t),
\label{eq:MCT-MSD-b}
\end{equation}
whereas the latter reads after carrying out the
$q \to 0$ limit in Eq.~(\ref{eq:delta-self})
\begin{equation}
\delta_{(0)}^{s}(z) = 
i \, w_{\rm hop} N_{\rm c} a^{2} / 6.
\label{eq:MCT-MSD-c}
\end{equation}
\end{subequations}
Equations~(\ref{eq:GLE-MSD}) and (\ref{eq:MCT-MSD}) constitute the
extended-MCT equations for the mean-squared displacement $\delta r^{2}(t)$. 

\subsection{Non-Gaussian parameter}
\label{appendix:NGP}

The non-Gaussian parameter $\alpha_{2}(t)$ 
defined in Eq.~(\ref{eq:NGP-def}) can be obtained from 
the mean-squared displacement 
$\delta r^{2}(t)$ and the mean-quartic displacement $\delta r^{4}(t)$.
The extended-MCT equations for $\delta r^{4}(t)$
can be derived using the same method employed 
above for $\delta r^{2}(t)$, but with higher order expansions in $q$. 

Since $\delta r^{4}(t)$
is proportional to the fourth Taylor coefficient 
in the small-$q$ expansion of 
$\phi_{q}^{s}(t) = 1 - q^{2} \delta r^{2}/3! +
q^{4} \delta r^{4} / 5! + O(q^{6})$~\cite{Hansen86},
one can derive the following equation from
the small-$q$ behavior 
of Eq.~(\ref{eq:GLE-phi-a}) for $\phi_{q}^{s}(t)$:
\begin{eqnarray}
& &
\partial_{t} \delta r^{4}(t) - 
20 (k_{\rm B}T/M) \int_{0}^{t} dt^{\prime} \, \delta r^{2}(t) 
\nonumber \\
& & 
\quad
+ \,
(k_{\rm B}T/M) \int_{0}^{t} dt^{\prime} \,
\Bigl[ \,
m_{(0)}^{s}(t-t^{\prime}) \, \delta r^{4}(t^{\prime}) 
\nonumber \\
& & 
\qquad \qquad \qquad 
- \, 
10 m_{(2)}^{s}(t-t^{\prime}) \, \delta r^{2}(t^{\prime})
\, \Bigr]  = 0.
\label{eq:GLE-MQD}
\end{eqnarray}
Here we introduced the memory kernel $m_{(2)}^{s}(t)$
via the small-$q$ expansion of 
$q^{2} m_{q}^{s}(t) =
m_{(0)}^{s}(t) + q^{2} m_{(2)}^{s}(t)/2 + O(q^{4})$.
Using the corresponding expansion for the idealized memory
kernel $m_{q}^{s \, {\rm id}}(t)$
and the expansion
$\delta_{q}^{s}(z) / q^{2} =
\delta_{(0)}^{s}(z) + q^{2} \delta_{(2)}^{s}(z)/2 + O(q^{4})$
for the hopping kernel,
one obtains from Eq.~(\ref{eq:extended-memory-for-appendix})
\begin{subequations}
\label{eq:MCT-MQD}
\begin{eqnarray}
m_{(2)}^{s}(z) &=&
m_{(2)}^{s \, {\rm id}}(z) + m_{(0)}^{s \, {\rm id}}(z) \, \delta_{(0)}^{s}(z) \, m_{(2)}^{s}(z)
\nonumber \\
& &
\quad 
+ \, 
m_{(0)}^{s \, {\rm id}}(z) \, \delta_{(2)}^{s}(z) \, m_{(0)}^{s}(z)
\nonumber \\
& &
\qquad
+ \, 
m_{(2)}^{s \, {\rm id}}(z) \, \delta_{(0)}^{s}(z) \, m_{(0)}^{s}(z).
\label{eq:MCT-MQD-a}
\end{eqnarray}
The expression for $m_{(2)}^{s \, {\rm id}}(t)$ reads~\cite{Fuchs98}
\begin{eqnarray}
m_{(2)}^{s \, {\rm id}}(t) &=&
\frac{1}{10 \pi^{2}}
\int dk \, k^{4} \rho S_{k} c_{k}^{2}
\phi_{k}(t) 
\nonumber \\
& & \qquad
\times \,
\left[
  \frac{\partial^{2} \phi_{k}^{s}(t)}{\partial k^{2}} +
   \frac{2}{3k} \frac{\partial \phi_{k}^{s}(t)}{\partial k}
\right],
\label{eq:MCT-MQD-b}
\end{eqnarray}
while one obtains from the small-$q$ expansion of Eq.~(\ref{eq:delta-self})
\begin{equation}
\delta_{(2)}^{s}(z) = i \, 
w_{\rm hop} N_{\rm c} a^{2} 
[ r_{s}^{2} - a^{2}/20] / 3. 
\label{eq:MCT-MQD-c}
\end{equation}
\end{subequations}
Here, $r_{s}$ is defined via the 
small-$q$ expansion of the Lamb-M\"ossbauer factor
$f_{q}^{s} = 1 - q^{2} r_{s}^{2} + O(q^{4})$~\cite{Fuchs98}.
Equations~(\ref{eq:GLE-MQD}) and (\ref{eq:MCT-MQD}) 
along with Eqs.~(\ref{eq:MCT-MSD}) constitute the
extended-MCT equations for $\delta r^{4}(t)$.
The non-Gaussian parameter $\alpha_{2}(t)$ can then be obtained from 
Eq.~(\ref{eq:NGP-def}).

Let us consider the short-time behavior of $\alpha_{2}(t)$
based on the derived equations. 
From Eqs.~(\ref{eq:GLE-MSD}) and (\ref{eq:GLE-MQD}) 
we find for short times
\begin{eqnarray}
\delta r^{2}(t) &=& 3 (k_{B}T/M)^{2} t^{2} + O(t^{4}),
\\
\delta r^{4}(t) &=& 5 (k_{B}T/M)^{4} 
\Bigl[ 1 + \frac{m_{(2)}^{s}(0)}{2} \Bigr] t^{4} + O(t^{6}).
\end{eqnarray}
Substituting these results into Eq.~(\ref{eq:NGP-def}), one obtains
\begin{equation}
\alpha_{2}(t) =
\Bigl[ \frac{m_{(2)}^{s}(0)}{6} - \frac{2}{3} \Bigr] + O(t^{2}).
\label{eq:NGP-short-times}
\end{equation}
Since $\phi_{q}^{s}(0) = 1$, 
we obtain from Eq.~(\ref{eq:MCT-MQD-b})
$m_{(2)}^{s \, {\rm id}}(0) = 0$.
Using this result, 
one can show that 
$m_{(2)}^{s}(0) = 0$ based on Eqs.~(\ref{eq:MCT-MQD-a}) and (\ref{eq:MCT-MQD-c}).
According to Eq.~(\ref{eq:NGP-short-times}), this means that
the initial value of $\alpha_{2}(t)$ within the extended MCT (and also within the
idealized MCT which is based on Newtonian dynamics) is given by
\begin{equation}
\alpha_{2}(t \to 0) = - 2/3.
\label{eq:NGP-short-times-MCT}
\end{equation}
This is in disagreement with the exact initial behavior $\alpha_{2}(t \to 0) = 0$~\cite{Hansen86}.
This discrepancy simply reflects that
the ideal-gas contribution to the memory kernel
is discarded in the mode-coupling approach. 
Indeed, the ideal-gas contribution yields
$m_{(2)}^{s}(0) = 4$~\cite{Hansen86}, which 
when substituted into Eq.~(\ref{eq:NGP-short-times}) 
recovers $\alpha_{2}(t \to 0) = 0$. 
The ideal-gas contribution to the memory kernel 
is responsible for the short-time ballistic regime,
but is irrelevant as far as the long-time dynamics is concerned.


\begin{thebibliography}{44}
\expandafter\ifx\csname natexlab\endcsname\relax\def\natexlab#1{#1}\fi
\expandafter\ifx\csname bibnamefont\endcsname\relax
  \def\bibnamefont#1{#1}\fi
\expandafter\ifx\csname bibfnamefont\endcsname\relax
  \def\bibfnamefont#1{#1}\fi
\expandafter\ifx\csname citenamefont\endcsname\relax
  \def\citenamefont#1{#1}\fi
\expandafter\ifx\csname url\endcsname\relax
  \def\url#1{\texttt{#1}}\fi
\expandafter\ifx\csname urlprefix\endcsname\relax\def\urlprefix{URL }\fi
\providecommand{\bibinfo}[2]{#2}
\providecommand{\eprint}[2][]{\url{#2}}

\bibitem[{\citenamefont{Chang and Sillescu}(1997)}]{Chang97}
\bibinfo{author}{\bibfnamefont{I.}~\bibnamefont{Chang}} \bibnamefont{and}
  \bibinfo{author}{\bibfnamefont{H.}~\bibnamefont{Sillescu}},
  \bibinfo{journal}{J. Phys. Chem. B} \textbf{\bibinfo{volume}{101}},
  \bibinfo{pages}{8794} (\bibinfo{year}{1997}).

\bibitem[{\citenamefont{Swallen et~al.}(2003)\citenamefont{Swallen, Bonvallet,
  {McMahon}, and Ediger}}]{Swallen03}
\bibinfo{author}{\bibfnamefont{S.~F.} \bibnamefont{Swallen}},
  \bibinfo{author}{\bibfnamefont{P.~A.} \bibnamefont{Bonvallet}},
  \bibinfo{author}{\bibfnamefont{R.~J.} \bibnamefont{{McMahon}}},
  \bibnamefont{and} \bibinfo{author}{\bibfnamefont{M.~D.}
  \bibnamefont{Ediger}}, \bibinfo{journal}{Phys. Rev. Lett.}
  \textbf{\bibinfo{volume}{90}}, \bibinfo{pages}{015901}
  (\bibinfo{year}{2003}).

\bibitem[{\citenamefont{Mapes et~al.}(2006)\citenamefont{Mapes, Swallen, and
  Ediger}}]{Mapes06}
\bibinfo{author}{\bibfnamefont{M.~K.} \bibnamefont{Mapes}},
  \bibinfo{author}{\bibfnamefont{S.~F.} \bibnamefont{Swallen}},
  \bibnamefont{and} \bibinfo{author}{\bibfnamefont{M.~D.}
  \bibnamefont{Ediger}}, \bibinfo{journal}{J. Phys. Chem. B}
  \textbf{\bibinfo{volume}{110}}, \bibinfo{pages}{507} (\bibinfo{year}{2006}).

\bibitem[{\citenamefont{Ediger}(2000)}]{Ediger00}
\bibinfo{author}{\bibfnamefont{M.~D.} \bibnamefont{Ediger}},
  \bibinfo{journal}{Annu. Rev. Phys. Chem.} \textbf{\bibinfo{volume}{51}},
  \bibinfo{pages}{99} (\bibinfo{year}{2000}).

\bibitem[{\citenamefont{Xia and Wolynes}(2001)}]{Xia01}
\bibinfo{author}{\bibfnamefont{X.}~\bibnamefont{Xia}} \bibnamefont{and}
  \bibinfo{author}{\bibfnamefont{P.~G.} \bibnamefont{Wolynes}},
  \bibinfo{journal}{J. Phys. Chem. B} \textbf{\bibinfo{volume}{105}},
  \bibinfo{pages}{6570} (\bibinfo{year}{2001}).

\bibitem[{\citenamefont{Berthier}(2004)}]{Berthier04}
\bibinfo{author}{\bibfnamefont{L.}~\bibnamefont{Berthier}},
  \bibinfo{journal}{Phys. Rev. E} \textbf{\bibinfo{volume}{69}},
  \bibinfo{pages}{020201(R)} (\bibinfo{year}{2004}).

\bibitem[{\citenamefont{Jung et~al.}(2004)\citenamefont{Jung, Garrahan, and
  Chandler}}]{Jung04}
\bibinfo{author}{\bibfnamefont{Y.~J.} \bibnamefont{Jung}},
  \bibinfo{author}{\bibfnamefont{J.~P.} \bibnamefont{Garrahan}},
  \bibnamefont{and} \bibinfo{author}{\bibfnamefont{D.}~\bibnamefont{Chandler}},
  \bibinfo{journal}{Phys. Rev. E} \textbf{\bibinfo{volume}{69}},
  \bibinfo{pages}{061205} (\bibinfo{year}{2004}).

\bibitem[{\citenamefont{R{\"o}ssler}(1990)}]{Roessler90}
\bibinfo{author}{\bibfnamefont{E.}~\bibnamefont{R{\"o}ssler}},
  \bibinfo{journal}{Ber. Bunsenges. Phys. Chem.} \textbf{\bibinfo{volume}{94}},
  \bibinfo{pages}{392} (\bibinfo{year}{1990}).

\bibitem[{\citenamefont{Goldstein}(1969)}]{Goldstein69}
\bibinfo{author}{\bibfnamefont{M.}~\bibnamefont{Goldstein}},
  \bibinfo{journal}{J. Chem. Phys.} \textbf{\bibinfo{volume}{51}},
  \bibinfo{pages}{3728} (\bibinfo{year}{1969}).

\bibitem[{\citenamefont{G{\"o}tze}(1991)}]{Goetze91b}
\bibinfo{author}{\bibfnamefont{W.}~\bibnamefont{G{\"o}tze}}, in
  \emph{\bibinfo{booktitle}{Liquids, Freezing and Glass Transition}}, edited by
  \bibinfo{editor}{\bibfnamefont{J.-P.} \bibnamefont{Hansen}},
  \bibinfo{editor}{\bibfnamefont{D.}~\bibnamefont{Levesque}}, \bibnamefont{and}
  \bibinfo{editor}{\bibfnamefont{J.}~\bibnamefont{Zinn-Justin}}
  (\bibinfo{publisher}{North-Holland}, \bibinfo{address}{Amsterdam},
  \bibinfo{year}{1991}), p. \bibinfo{pages}{287}.

\bibitem[{\citenamefont{G{\"o}tze and Sj{\"o}gren}(1992)}]{Goetze92}
\bibinfo{author}{\bibfnamefont{W.}~\bibnamefont{G{\"o}tze}} \bibnamefont{and}
  \bibinfo{author}{\bibfnamefont{L.}~\bibnamefont{Sj{\"o}gren}},
  \bibinfo{journal}{Rep. Prog. Phys.} \textbf{\bibinfo{volume}{55}},
  \bibinfo{pages}{241} (\bibinfo{year}{1992}).

\bibitem[{\citenamefont{G{\"o}tze}(1999)}]{Goetze99}
\bibinfo{author}{\bibfnamefont{W.}~\bibnamefont{G{\"o}tze}},
  \bibinfo{journal}{J. Phys.: Condensed Matter} \textbf{\bibinfo{volume}{11}},
  \bibinfo{pages}{A1} (\bibinfo{year}{1999}).

\bibitem[{\citenamefont{G{\"o}tze and Sj{\"o}gren}(1987)}]{Goetze87}
\bibinfo{author}{\bibfnamefont{W.}~\bibnamefont{G{\"o}tze}} \bibnamefont{and}
  \bibinfo{author}{\bibfnamefont{L.}~\bibnamefont{Sj{\"o}gren}},
  \bibinfo{journal}{Z. Phys. B} \textbf{\bibinfo{volume}{65}},
  \bibinfo{pages}{415} (\bibinfo{year}{1987}).

\bibitem[{\citenamefont{Kawasaki}(1994)}]{Kawasaki94}
\bibinfo{author}{\bibfnamefont{K.}~\bibnamefont{Kawasaki}},
  \bibinfo{journal}{Physica A} \textbf{\bibinfo{volume}{208}},
  \bibinfo{pages}{35} (\bibinfo{year}{1994}).

\bibitem[{\citenamefont{Schweizer and Saltzman}(2003)}]{Schweizer03}
\bibinfo{author}{\bibfnamefont{K.~S.} \bibnamefont{Schweizer}}
  \bibnamefont{and} \bibinfo{author}{\bibfnamefont{E.~J.}
  \bibnamefont{Saltzman}}, \bibinfo{journal}{J. Chem. Phys.}
  \textbf{\bibinfo{volume}{119}}, \bibinfo{pages}{1181} (\bibinfo{year}{2003}).

\bibitem[{\citenamefont{Bhattacharyya et~al.}(2005)\citenamefont{Bhattacharyya,
  Bagchi, and Wolynes}}]{Bhattacharyya05}
\bibinfo{author}{\bibfnamefont{S.~M.} \bibnamefont{Bhattacharyya}},
  \bibinfo{author}{\bibfnamefont{B.}~\bibnamefont{Bagchi}}, \bibnamefont{and}
  \bibinfo{author}{\bibfnamefont{P.~G.} \bibnamefont{Wolynes}},
  \bibinfo{journal}{Phys. Rev. E} \textbf{\bibinfo{volume}{72}},
  \bibinfo{pages}{031509} (\bibinfo{year}{2005}).

\bibitem[{Fly()}]{Flynn-all}
\bibinfo{note}{C.~P.~Flynn, Phys.~Rev.~{\bf 171}, 682 (1968); C.~P.~Flynn, {\em
  Point defects and diffusion} (Clarendon Press, Oxford, 1972).}

\bibitem[{\citenamefont{Z{\"o}llmer et~al.}(2003)\citenamefont{Z{\"o}llmer,
  R{\"a}tzke, Faupel, and Meyer}}]{Zollmer03}
\bibinfo{author}{\bibfnamefont{V.}~\bibnamefont{Z{\"o}llmer}},
  \bibinfo{author}{\bibfnamefont{K.}~\bibnamefont{R{\"a}tzke}},
  \bibinfo{author}{\bibfnamefont{F.}~\bibnamefont{Faupel}}, \bibnamefont{and}
  \bibinfo{author}{\bibfnamefont{A.}~\bibnamefont{Meyer}},
  \bibinfo{journal}{Phys. Rev. Lett.} \textbf{\bibinfo{volume}{90}},
  \bibinfo{pages}{195502} (\bibinfo{year}{2003}).

\bibitem[{\citenamefont{Kob et~al.}(1997)\citenamefont{Kob, Donati, Plimpton,
  Poole, and Glotzer}}]{Kob97}
\bibinfo{author}{\bibfnamefont{W.}~\bibnamefont{Kob}},
  \bibinfo{author}{\bibfnamefont{C.}~\bibnamefont{Donati}},
  \bibinfo{author}{\bibfnamefont{S.~J.} \bibnamefont{Plimpton}},
  \bibinfo{author}{\bibfnamefont{P.~H.} \bibnamefont{Poole}}, \bibnamefont{and}
  \bibinfo{author}{\bibfnamefont{S.~C.} \bibnamefont{Glotzer}},
  \bibinfo{journal}{Phys. Rev. Lett.} \textbf{\bibinfo{volume}{79}},
  \bibinfo{pages}{2827} (\bibinfo{year}{1997}).

\bibitem[{\citenamefont{Flenner and Szamel}(2005{\natexlab{a}})}]{Flenner05}
\bibinfo{author}{\bibfnamefont{E.}~\bibnamefont{Flenner}} \bibnamefont{and}
  \bibinfo{author}{\bibfnamefont{G.}~\bibnamefont{Szamel}},
  \bibinfo{journal}{Phys. Rev. E} \textbf{\bibinfo{volume}{72}},
  \bibinfo{pages}{011205} (\bibinfo{year}{2005}{\natexlab{a}}).

\bibitem[{\citenamefont{Flenner and Szamel}(2005{\natexlab{b}})}]{Flenner05b}
\bibinfo{author}{\bibfnamefont{E.}~\bibnamefont{Flenner}} \bibnamefont{and}
  \bibinfo{author}{\bibfnamefont{G.}~\bibnamefont{Szamel}},
  \bibinfo{journal}{Phys. Rev. E} \textbf{\bibinfo{volume}{72}},
  \bibinfo{pages}{031508} (\bibinfo{year}{2005}{\natexlab{b}}).

\bibitem[{\citenamefont{Hansen and McDonald}(1986)}]{Hansen86}
\bibinfo{author}{\bibfnamefont{J.-P.} \bibnamefont{Hansen}} \bibnamefont{and}
  \bibinfo{author}{\bibfnamefont{I.~R.} \bibnamefont{McDonald}},
  \emph{\bibinfo{title}{Theory of Simple Liquids}}
  (\bibinfo{publisher}{Academic Press}, \bibinfo{address}{London},
  \bibinfo{year}{1986}), \bibinfo{edition}{2nd} ed.

\bibitem[{\citenamefont{G{\"o}tze and Sj{\"o}gren}(1995)}]{Goetze95}
\bibinfo{author}{\bibfnamefont{W.}~\bibnamefont{G{\"o}tze}} \bibnamefont{and}
  \bibinfo{author}{\bibfnamefont{L.}~\bibnamefont{Sj{\"o}gren}},
  \bibinfo{journal}{Transp. Theory Stat. Phys.} \textbf{\bibinfo{volume}{24}},
  \bibinfo{pages}{801} (\bibinfo{year}{1995}).

\bibitem[{\citenamefont{Roux et~al.}(1989)\citenamefont{Roux, Barrat, and
  Hansen}}]{Roux89}
\bibinfo{author}{\bibfnamefont{J.~N.} \bibnamefont{Roux}},
  \bibinfo{author}{\bibfnamefont{J.-L.} \bibnamefont{Barrat}},
  \bibnamefont{and} \bibinfo{author}{\bibfnamefont{J.-P.}
  \bibnamefont{Hansen}}, \bibinfo{journal}{J. Phys.: Condens. Matter}
  \textbf{\bibinfo{volume}{1}}, \bibinfo{pages}{7171} (\bibinfo{year}{1989}).

\bibitem[{\citenamefont{Wahnstr{\"o}m}(1991)}]{Wahnstroem91}
\bibinfo{author}{\bibfnamefont{G.}~\bibnamefont{Wahnstr{\"o}m}},
  \bibinfo{journal}{Phys. Rev. A} \textbf{\bibinfo{volume}{44}},
  \bibinfo{pages}{3752} (\bibinfo{year}{1991}).

\bibitem[{\citenamefont{Sch{\o}der et~al.}(2000)\citenamefont{Sch{\o}der,
  Sastry, Dyre, and Glotzer}}]{Schroder00}
\bibinfo{author}{\bibfnamefont{T.~B.} \bibnamefont{Sch{\o}der}},
  \bibinfo{author}{\bibfnamefont{S.}~\bibnamefont{Sastry}},
  \bibinfo{author}{\bibfnamefont{J.~C.} \bibnamefont{Dyre}}, \bibnamefont{and}
  \bibinfo{author}{\bibfnamefont{S.~C.} \bibnamefont{Glotzer}},
  \bibinfo{journal}{J. Chem. Phys.} \textbf{\bibinfo{volume}{112}},
  \bibinfo{pages}{9834} (\bibinfo{year}{2000}).

\bibitem[{\citenamefont{Barrat et~al.}(1989)\citenamefont{Barrat, G{\"o}tze,
  and Latz}}]{Barrat89}
\bibinfo{author}{\bibfnamefont{J.-L.} \bibnamefont{Barrat}},
  \bibinfo{author}{\bibfnamefont{W.}~\bibnamefont{G{\"o}tze}},
  \bibnamefont{and} \bibinfo{author}{\bibfnamefont{A.}~\bibnamefont{Latz}},
  \bibinfo{journal}{J. Phys.: Condens. Matter} \textbf{\bibinfo{volume}{1}},
  \bibinfo{pages}{7163} (\bibinfo{year}{1989}).

\bibitem[{\citenamefont{G{\"o}tze and Mayr}(2000)}]{Goetze00}
\bibinfo{author}{\bibfnamefont{W.}~\bibnamefont{G{\"o}tze}} \bibnamefont{and}
  \bibinfo{author}{\bibfnamefont{M.~R.} \bibnamefont{Mayr}},
  \bibinfo{journal}{Phys. Rev. E} \textbf{\bibinfo{volume}{61}},
  \bibinfo{pages}{587} (\bibinfo{year}{2000}).

\bibitem[{\citenamefont{Chong}(2006)}]{Chong06}
\bibinfo{author}{\bibfnamefont{S.-H.} \bibnamefont{Chong}},
  \bibinfo{journal}{Phys. Rev. E} \textbf{\bibinfo{volume}{74}},
  \bibinfo{pages}{031205} (\bibinfo{year}{2006}).

\bibitem[{\citenamefont{G{\"o}tze and Sperl}(2003)}]{Goetze03}
\bibinfo{author}{\bibfnamefont{W.}~\bibnamefont{G{\"o}tze}} \bibnamefont{and}
  \bibinfo{author}{\bibfnamefont{M.}~\bibnamefont{Sperl}}, \bibinfo{journal}{J.
  Phys.: Condens. Matter} \textbf{\bibinfo{volume}{15}}, \bibinfo{pages}{S869}
  (\bibinfo{year}{2003}).

\bibitem[{\citenamefont{Fuchs et~al.}(1998)\citenamefont{Fuchs, G{\"o}tze, and
  Mayr}}]{Fuchs98}
\bibinfo{author}{\bibfnamefont{M.}~\bibnamefont{Fuchs}},
  \bibinfo{author}{\bibfnamefont{W.}~\bibnamefont{G{\"o}tze}},
  \bibnamefont{and} \bibinfo{author}{\bibfnamefont{M.~R.} \bibnamefont{Mayr}},
  \bibinfo{journal}{Phys. Rev. E} \textbf{\bibinfo{volume}{58}},
  \bibinfo{pages}{3384} (\bibinfo{year}{1998}).

\bibitem[{com({\natexlab{a}})}]{comment-Lindemann-length}
\bibinfo{note}{Assuming $s \approx a/2$ and $x^{*} \approx \lim_{t \to \infty}
  \sqrt{ \delta r^{2}(t) }$ evaluated within the idealized MCT at the critical
  point, which corresponds to the Lindemann length~\cite{Fuchs98}, one obtains
  $\Delta^{2} \approx 0.12$ for the Lennard-Jones system under study and
  $\Delta^{2} \approx 0.10$ for the hard-sphere system studied in
  Ref.~\cite{Fuchs98}.}

\bibitem[{\citenamefont{Franosch et~al.}(1997)\citenamefont{Franosch, Fuchs,
  G{\"o}tze, Mayr, and Singh}}]{Franosch97}
\bibinfo{author}{\bibfnamefont{T.}~\bibnamefont{Franosch}},
  \bibinfo{author}{\bibfnamefont{M.}~\bibnamefont{Fuchs}},
  \bibinfo{author}{\bibfnamefont{W.}~\bibnamefont{G{\"o}tze}},
  \bibinfo{author}{\bibfnamefont{M.~R.} \bibnamefont{Mayr}}, \bibnamefont{and}
  \bibinfo{author}{\bibfnamefont{A.~P.} \bibnamefont{Singh}},
  \bibinfo{journal}{Phys. Rev. E} \textbf{\bibinfo{volume}{55}},
  \bibinfo{pages}{7153} (\bibinfo{year}{1997}).

\bibitem[{\citenamefont{Mezei et~al.}(1987)\citenamefont{Mezei, Knaak, and
  Farago}}]{Mezei87}
\bibinfo{author}{\bibfnamefont{F.}~\bibnamefont{Mezei}},
  \bibinfo{author}{\bibfnamefont{W.}~\bibnamefont{Knaak}}, \bibnamefont{and}
  \bibinfo{author}{\bibfnamefont{B.}~\bibnamefont{Farago}},
  \bibinfo{journal}{Phys. Rev. Lett.} \textbf{\bibinfo{volume}{58}},
  \bibinfo{pages}{571} (\bibinfo{year}{1987}).

\bibitem[{\citenamefont{Yamamoto and Onuki}(1998)}]{Yamamoto98}
\bibinfo{author}{\bibfnamefont{R.}~\bibnamefont{Yamamoto}} \bibnamefont{and}
  \bibinfo{author}{\bibfnamefont{A.}~\bibnamefont{Onuki}},
  \bibinfo{journal}{Phys. Rev. Lett.} \textbf{\bibinfo{volume}{81}},
  \bibinfo{pages}{4915} (\bibinfo{year}{1998}).

\bibitem[{\citenamefont{Li et~al.}(1992)\citenamefont{Li, Du, Sakai, and
  Cummins}}]{Li92b}
\bibinfo{author}{\bibfnamefont{G.}~\bibnamefont{Li}},
  \bibinfo{author}{\bibfnamefont{W.~M.} \bibnamefont{Du}},
  \bibinfo{author}{\bibfnamefont{A.}~\bibnamefont{Sakai}}, \bibnamefont{and}
  \bibinfo{author}{\bibfnamefont{H.~Z.} \bibnamefont{Cummins}},
  \bibinfo{journal}{Phys. Rev. A} \textbf{\bibinfo{volume}{46}},
  \bibinfo{pages}{3343} (\bibinfo{year}{1992}).

\bibitem[{\citenamefont{Kob and Andersen}(1994)}]{Kob94}
\bibinfo{author}{\bibfnamefont{W.}~\bibnamefont{Kob}} \bibnamefont{and}
  \bibinfo{author}{\bibfnamefont{H.~C.} \bibnamefont{Andersen}},
  \bibinfo{journal}{Phys. Rev. Lett.} \textbf{\bibinfo{volume}{73}},
  \bibinfo{pages}{1376} (\bibinfo{year}{1994}).

\bibitem[{\citenamefont{Chong et~al.}(2005)\citenamefont{Chong, Moreno,
  Sciortino, and Kob}}]{Chong05}
\bibinfo{author}{\bibfnamefont{S.-H.} \bibnamefont{Chong}},
  \bibinfo{author}{\bibfnamefont{A.~J.} \bibnamefont{Moreno}},
  \bibinfo{author}{\bibfnamefont{F.}~\bibnamefont{Sciortino}},
  \bibnamefont{and} \bibinfo{author}{\bibfnamefont{W.}~\bibnamefont{Kob}},
  \bibinfo{journal}{Phys. Rev. Lett.} \textbf{\bibinfo{volume}{94}},
  \bibinfo{pages}{215701} (\bibinfo{year}{2005}).

\bibitem[{com({\natexlab{b}})}]{comment-comparison-with-simulation}
\bibinfo{note}{Simulation results for the product $D \tau_{q^{*}}$ are formed
  with the $\alpha$-relaxation time of the tagged-particle density correlator
  rather than that of the coherent density correlator since the statistics is
  much better for the former. The ratios $R$ from the simulations are then
  calculated by rescaling the product by the one at a reference temperature
  $T_{\rm ref}$ and are plotted versus $\epsilon = \tilde{C} (T_{\rm c} - T) /
  T_{\rm c}$ with $T_{\rm ref} = 0.80$, $\tilde{C} = 0.12$, and $T_{\rm c} =
  0.435$ for the binary mixture of Lennard-Jones particles~\cite{Kob94}, and
  with $T_{\rm ref} = 3.0$, $\tilde{C} = 0.17$, and $T_{\rm c} = 2.10$ for the
  binary mixture of dumbbell molecules of elongation $\zeta =
  0.8$~\cite{Chong05}. Here $\tilde{C}$ refers to the ratio $C_{\rm MD} /
  C_{\rm LJ}$, where $C_{\rm MD}$ and $C_{\rm LJ}$ are the constants $C$ for
  the simulated systems and for our Lennard-Jones system, respectively,
  connecting the separation parameter $\sigma$ relevant in MCT and the reduced
  temperature $\epsilon = (T_{\rm c} - T)/T_{\rm c}$ via $\sigma = C
  \epsilon$~\cite{Goetze91b}. For the comparison shown in the inset of
  Fig.~\ref{fig:Dtau}, it was necessary to rescale the reduced temperatures of
  the simulated systems by $\tilde{C}$ to absorb difference in $C_{\rm MD}$ and
  $C_{\rm LJ}$.}

\bibitem[{\citenamefont{Rahman}(1964)}]{Rahman64}
\bibinfo{author}{\bibfnamefont{A.}~\bibnamefont{Rahman}},
  \bibinfo{journal}{Phys. Rev. A} \textbf{\bibinfo{volume}{136}},
  \bibinfo{pages}{405} (\bibinfo{year}{1964}).

\bibitem[{\citenamefont{La\v{c}evi\'{c}
  et~al.}(2003)\citenamefont{La\v{c}evi\'{c}, Starr, Sch{\o}der, and
  Glotzer}}]{Lacevic03}
\bibinfo{author}{\bibfnamefont{N.}~\bibnamefont{La\v{c}evi\'{c}}},
  \bibinfo{author}{\bibfnamefont{F.~W.} \bibnamefont{Starr}},
  \bibinfo{author}{\bibfnamefont{T.~B.} \bibnamefont{Sch{\o}der}},
  \bibnamefont{and} \bibinfo{author}{\bibfnamefont{S.~C.}
  \bibnamefont{Glotzer}}, \bibinfo{journal}{J. Chem. Phys.}
  \textbf{\bibinfo{volume}{119}}, \bibinfo{pages}{7372} (\bibinfo{year}{2003}).

\bibitem[{\citenamefont{Donati et~al.}(1998)\citenamefont{Donati, Douglas, Kob,
  Plimpton, Poole, and Glotzer}}]{Donati98}
\bibinfo{author}{\bibfnamefont{C.}~\bibnamefont{Donati}},
  \bibinfo{author}{\bibfnamefont{J.~F.} \bibnamefont{Douglas}},
  \bibinfo{author}{\bibfnamefont{W.}~\bibnamefont{Kob}},
  \bibinfo{author}{\bibfnamefont{S.~J.} \bibnamefont{Plimpton}},
  \bibinfo{author}{\bibfnamefont{P.~H.} \bibnamefont{Poole}}, \bibnamefont{and}
  \bibinfo{author}{\bibfnamefont{S.~C.} \bibnamefont{Glotzer}},
  \bibinfo{journal}{Phys. Rev. Lett.} \textbf{\bibinfo{volume}{80}},
  \bibinfo{pages}{2338} (\bibinfo{year}{1998}).

\bibitem[{\citenamefont{Feynman and Cohen}(1956)}]{Feynman56}
\bibinfo{author}{\bibfnamefont{R.~P.} \bibnamefont{Feynman}} \bibnamefont{and}
  \bibinfo{author}{\bibfnamefont{M.}~\bibnamefont{Cohen}},
  \bibinfo{journal}{Phys. Rev.} \textbf{\bibinfo{volume}{102}},
  \bibinfo{pages}{1189} (\bibinfo{year}{1956}).

\bibitem[{\citenamefont{G{\"o}tze and L{\"u}cke}(1976)}]{Goetze76}
\bibinfo{author}{\bibfnamefont{W.}~\bibnamefont{G{\"o}tze}} \bibnamefont{and}
  \bibinfo{author}{\bibfnamefont{M.}~\bibnamefont{L{\"u}cke}},
  \bibinfo{journal}{Phys. Rev. B} \textbf{\bibinfo{volume}{13}},
  \bibinfo{pages}{3825} (\bibinfo{year}{1976}).

\end{thebibliography}
\end{document}